\documentclass[11pt,twoside,english]{elsarticle}
\usepackage[T1]{fontenc}
\usepackage[latin9]{inputenc}
\usepackage{textcomp}
\usepackage{mathtools}
\usepackage{amsmath}
\usepackage{graphicx}
\usepackage{hyperref}
\hypersetup{
colorlinks=true,
linkcolor=blue,
filecolor=magenta,
urlcolor=blue
}

\makeatletter

\newcommand{\lyxmathsym}[1]{\ifmmode\begingroup\def\b@ld{bold}
  \text{\ifx\math@version\b@ld\bfseries\fi#1}\endgroup\else#1\fi}

\providecommand{\tabularnewline}{\\}

\journal{Acta Materialia}
\biboptions{sort&compress}
\usepackage{geometry}
\@ifundefined{showcaptionsetup}{}{%
 \PassOptionsToPackage{caption=false}{subfig}}
\usepackage{subfig}
\makeatother
\usepackage{babel}
\usepackage{xcolor}
\begin{document}

\begin{frontmatter}{}

\title{Phase-field thermo-electromechanical modelling of lead-free BNT-based
piezoelectric materials}

\author[rvt]{Akshayveer\corref{cor1}}

\ead{aakshayveer@wlu.ca}

\author[focal]{Federico C.~Buroni }

\author[rvt]{Roderick Melnik }

\author[els]{Luis ~Rodriguez-Tembleque }

\author[els]{Andres ~Saez }

\cortext[cor1]{Corresponding author}

\address[rvt]{MS2Discovery Interdisciplinary Research Institute, Wilfrid Laurier
University, Waterloo, Ontario N2L3C5, Canada }

\address[focal]{Department of Mechanical Engineering and Manufacturing, Universidad
de Sevilla, Camino de los Descubrimientos s/n, Seville E-41092, Spain}

\address[els]{Department of Continuum Mechanics and Structural Analysis, Universidad
de Sevilla, Camino de los Descubrimientos s/n, Seville E-41092, Spain}
\begin{abstract}
In recent years, bismuth sodium titanate (BNT) and BNT-based piezoelectric materials have emerged as promising candidates for lead-free piezoelectric technology. However,
these materials possess a complex phase structure that plays a vital
role in their thermo-electromechanical behaviour. With the evolution
of thermal conditions, the thermo-electromechanical response alters
due to phase change and micro-domain switching. The pure BNT material
has a rhombohedral (R3c) lattice structure at ambient temperature.
However, it undergoes a phase transition to an orthorhombic phase
(Pnma) at 200\textcelsius, mostly owing to octahedral tilting. This transition
is known as the depolarization temperature ($T_{d}$). The ferroelectric
domains in the R3c phase have a larger spontaneous polarization, which
decreases as temperature rises and reaches a local minimum at $T_{d}$,
leading to a phase transition. Moreover, BNT undergoes a transition
from the Pnma phase to the tetragonal phase (P4bm) at a temperature
of 320\textcelsius, and from the P4bm phase to the cubic phase (Pm3m) at a temperature
of 520\textcelsius. These transitions also result in the changes in the spontaneous
polarization. Therefore, the investigation of domain switching and
phase change dynamics in relation to temperature and electromechanical
coupling has acquired considerable importance. The current work aims
to assess the complex phase regimes and their activation under various
thermal environments to enhance the utilization of these materials
for specific applications, including sensors, actuators, energy harvesting
devices, and haptic technologies. The micro-sized BNT inclusions are embedded with polymer matrix such as polydimethylsiloxane (PDMS), which represents a practical scenario. A two-dimensional phase-field thermo-electromechanical
computational model has been created to investigate the complex phase
transition and domain switching behaviour of BNT-PDMS composite under varying
temperature conditions. The model includes Landau-Ginzburg free energy
and thermo-electromechanical free energy to effectively represent
the impact of phase shift and domain switching. The model has been
validated with experimental data for pure BNT. It can accurately forecast
the thermo-electromechanical response in different phase regimes and
the coexistence of phases due to temperature changes. 
\end{abstract}

\end{frontmatter}{}

\section{Introduction}

\begin{figure}
\subfloat[]{\includegraphics[scale=0.5]{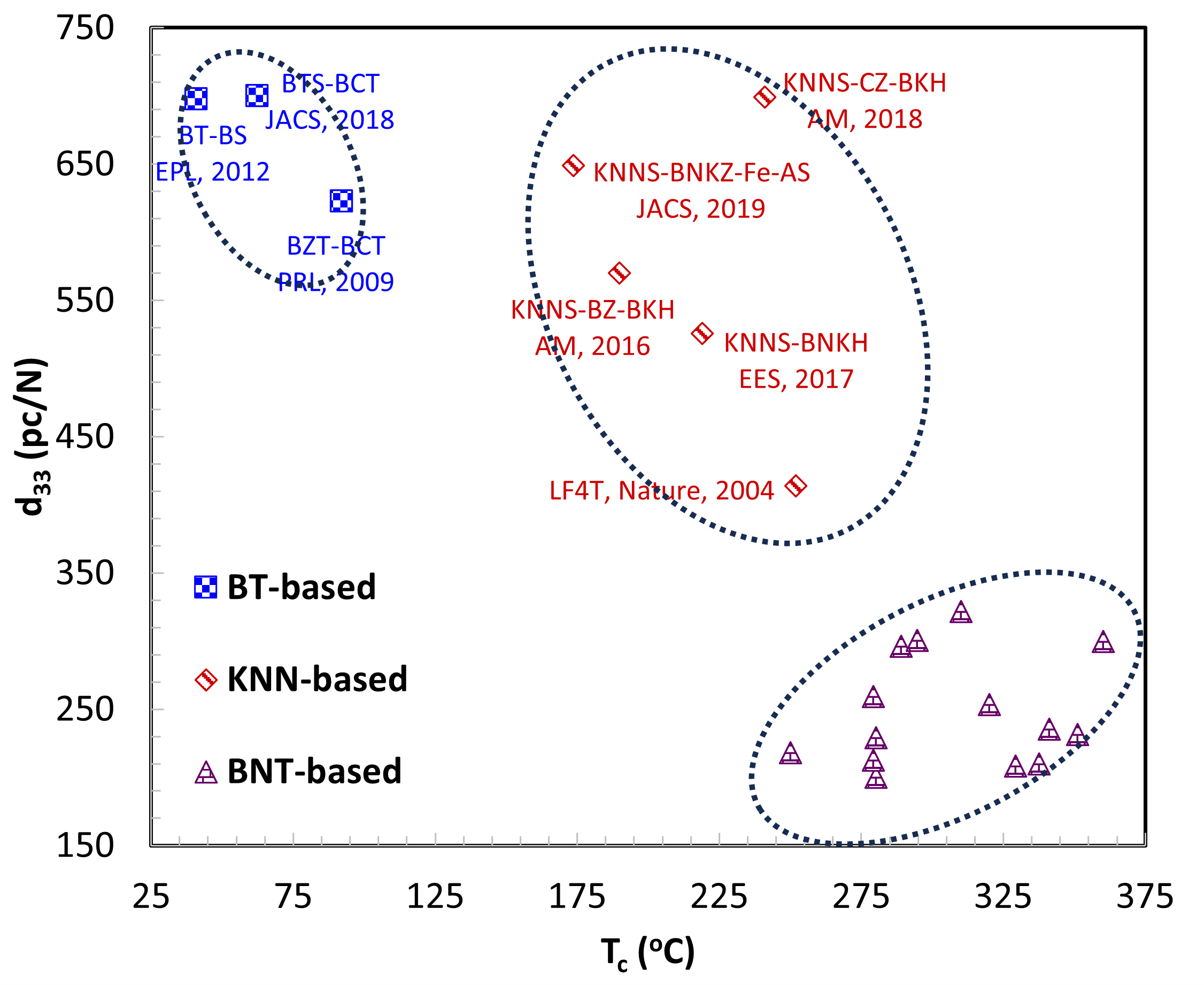}

}\subfloat[]{\includegraphics[scale=0.5]{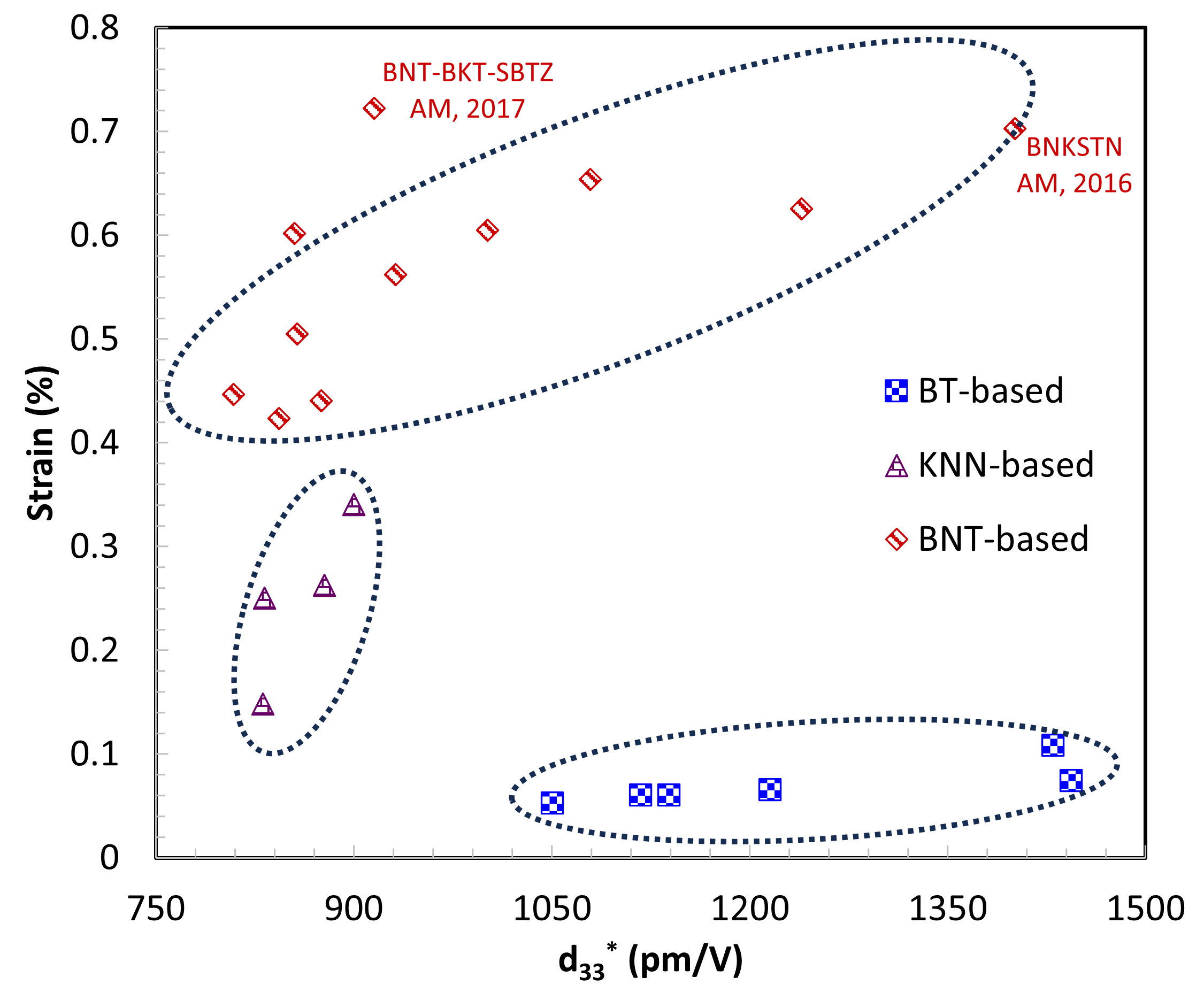}

}

\caption{Schematic diagram showing (a) piezoelectric response $d_{33}$, and
(b) strain for various BT, KNN, BNT-based piezoelectric composites. Reproduced with permission from Ref.
\citep{Zhou2021}. Copyright 2021 Elsevier. \label{fig:Schematic-diagram-showing}}
\end{figure}

Lead-free piezoelectric composites offer scalable and eco-friendly
options for electromechanical actuators and mechanical energy harvesting
and sensing \citep{article,Maurya2018,Tiller2019}. These composites
typically consist of relatively softer matrices, typically polymers,
filled with nano- or micro-particles of hard, crystalline, lead-free
piezoelectric materials with vigorous piezoelectric activity, such
as non-perovskites like bismuth layered structured ferroelectrics
(BLSF) and tungsten bronze ferroelectrics \citep{Takenaka2005}, and
perovskites like BaTiO$_{3}$ (BT), Bi$_{0.5}$Na$_{0.5}$TiO$_{3}$ (BNT),
and KNaNbO$_{3}$ (KNN) \citep{Shin2014,Quan2021,Zhang2023}. These
lead-free piezoelectric materials have lower piezoelectric coefficients,
but their piezoelectric performance may be enhanced by intrinsic (lattice
distortions) and extrinsic contributions (altering the domain topologies)
\citep{Damjanovic1998}. The novel piezoelectric materials based on
KNN and BT demonstrate enhanced piezoelectric response ($d_{33}\sim500pc/N$)
and ($d_{33}\sim445\pm20pc/N$) respectively through the manipulation
of lattice softening, and reduction in unit cell distortion, however,
the low Curie temperature ($T_{c}$) for KNN-based composites ($T_{c}\sim200\lyxmathsym{\textcelsius}$)
and BT-based composites ($T_{c}<100\lyxmathsym{\textcelsius}$) limits
their usage for high temperature haptic and actuator applications\citep{Liu2019,Wang2021}.
Pure BNT has depicted moderate piezoelectric properties $(d_{33}<100pc/N)$,
and high coercive electric field $(E_{c}\sim70kV/cm)$ \citep{Hao2019}
but BNT also exhibit high value of $T_{c}$ (320\textcelsius ) \citep{Kumari2022}
which enhances its adaptability to high temperature haptic applications.
Moreover, BNT-based composites exhibit more strain than KNN and BT-based
composites in high temperature regimes (above 200\textcelsius ) \citep{Zhou2021},
which also make it more suitable for high temperature actuator applications.
A state of the art of various BT, KNN, and BNT-based composites is
shown in Fig. \ref{fig:Schematic-diagram-showing} which clearly supports
the fact that BNT has the functionality to work in higher temperature
zones, and moderate piezoelectric response and high strain make it
more suitable for high temperature sensor, haptic and actuator based
applications. Therefore, we have chosen BNT and BNT-based composites
for the current study due to its high adaptibility for high temperature
haptic applications despite the fact that BNT exhibits a more complex
phase structure than other group of lead-free piezoelectric composites.
BNT is a ferroelectric with an A-site disordered relaxor \citep{Smolensky1961}.
The hybridization between the Bi 6s and O 2p orbitals is responsible
for the substantial spontaneous polarization of around 40 $\mu C/cm^{2}$
observed in BNT at ambient temperature \citep{JAFFE1958}. With increasing
temperature, the dielectric-temperature spectra show two diffused
dielectric peaks: a frequency-dependent hump at 200 $\lyxmathsym{\textcelsius}$
and a broad maximum at 320 $\lyxmathsym{\textcelsius}$ $(T_{m})$
\citep{Tu1994}. Furthermore, a decline of spontaneous polarization
can be observed beyond 200 $\lyxmathsym{\textcelsius}$, which is
characterized as the depolarization temperature ($T_{d}$). 

Many research works \citep{Tu1994,Rao2013,Jones2002,Aksel2011,Aksel2013,Gorfman2010,Rao2013a,Trolliard2008,Dorcet2009,Dorcet2008,Dorcet2008a,GOMAHPETTRY2011,Suchanicz1995,J.Suchanicz1988,J.A.Zvirgzds1982,Jones2000}
have been conducted to give fresh light on the complex phase structures
and temperature dependent phase transitions of BNT, however some disagreements
remain. Depending on thermal, electrical, and mechanical treatments,
the structure has been reported to be rhombohedral (R3c) \citep{Rao2013,Jones2002},
monoclinic (Cc) \citep{Aksel2011,Aksel2013}, or a mixture of both
\citep{Gorfman2010,Rao2013a} below 200$\lyxmathsym{\textcelsius}$.
BNT exhibits a ferroelectric tetragonal (P4bm) phase above 320$\lyxmathsym{\textcelsius}$
\citep{Jones2002,Jones2000}. Some investigations \citep{Trolliard2008,Dorcet2009,Dorcet2008,Dorcet2008a}
have reported an intermediate orthorhombic (Pnma) phase with an antiferroelectric
(AFE) character between 200-320 $\lyxmathsym{\textcelsius}$, however
the presence of an AFE phase is being questioned, and most researchers
now believe it to be a nonpolar or weakly-polar phase \citep{Rao2013,GOMAHPETTRY2011,Suchanicz1995,J.Suchanicz1988}.
Many elements of BNT crystal structures are still unknown, particularly
octahedral tilting \citep{Jones2002,Jones2000} and A-site cation
displacement behaviors \citep{Aksel2013,Kreisel2003,Shuvaeva2005,Keeble2013},
which have a significant impact on the phase, polarization, and relaxor
ferroelectric properties of BNT. 

\begin{figure}
\subfloat[]{\includegraphics[width=3.2in,height=3in]{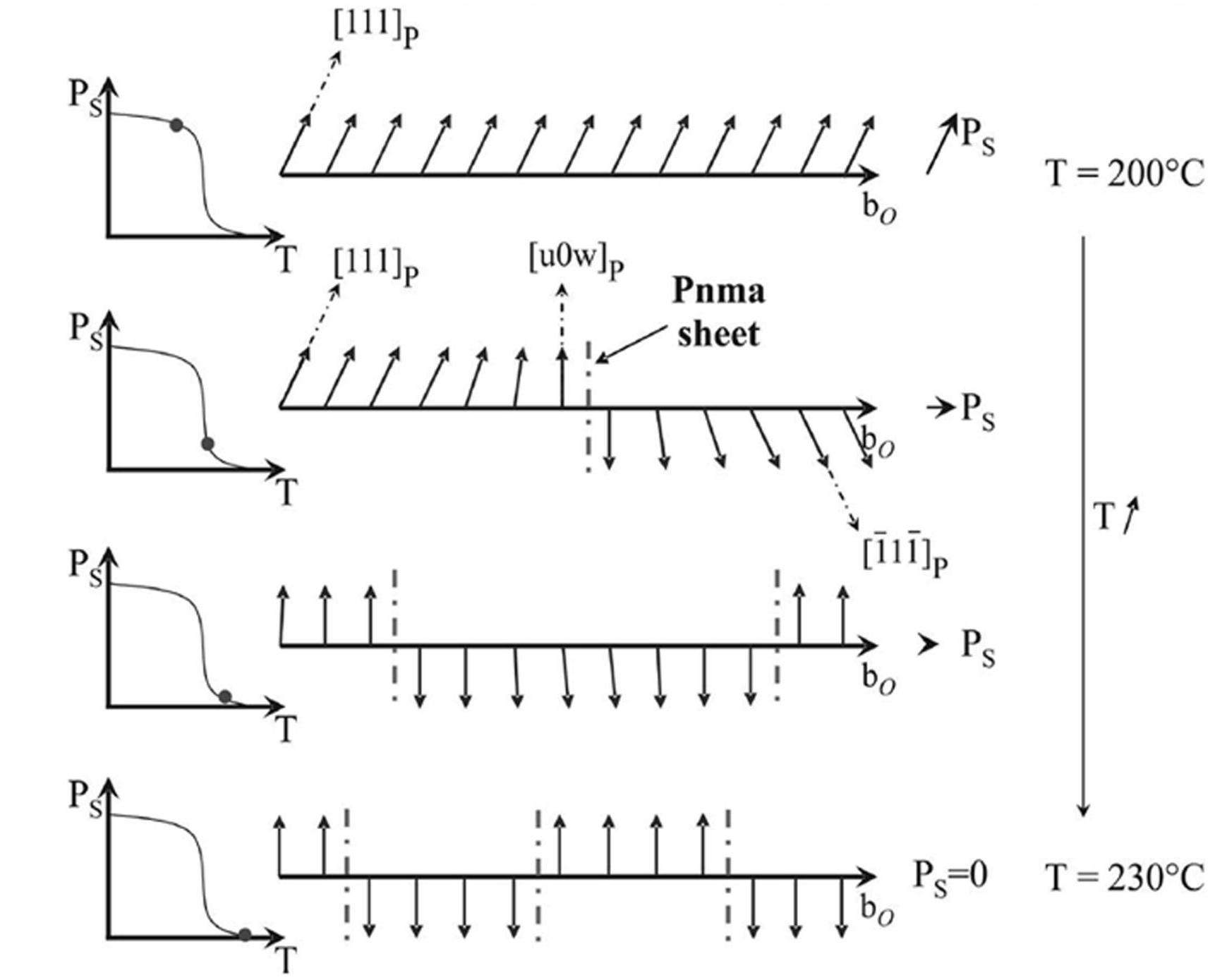}

}\subfloat[]{\includegraphics[width=3.2in,height=3in]{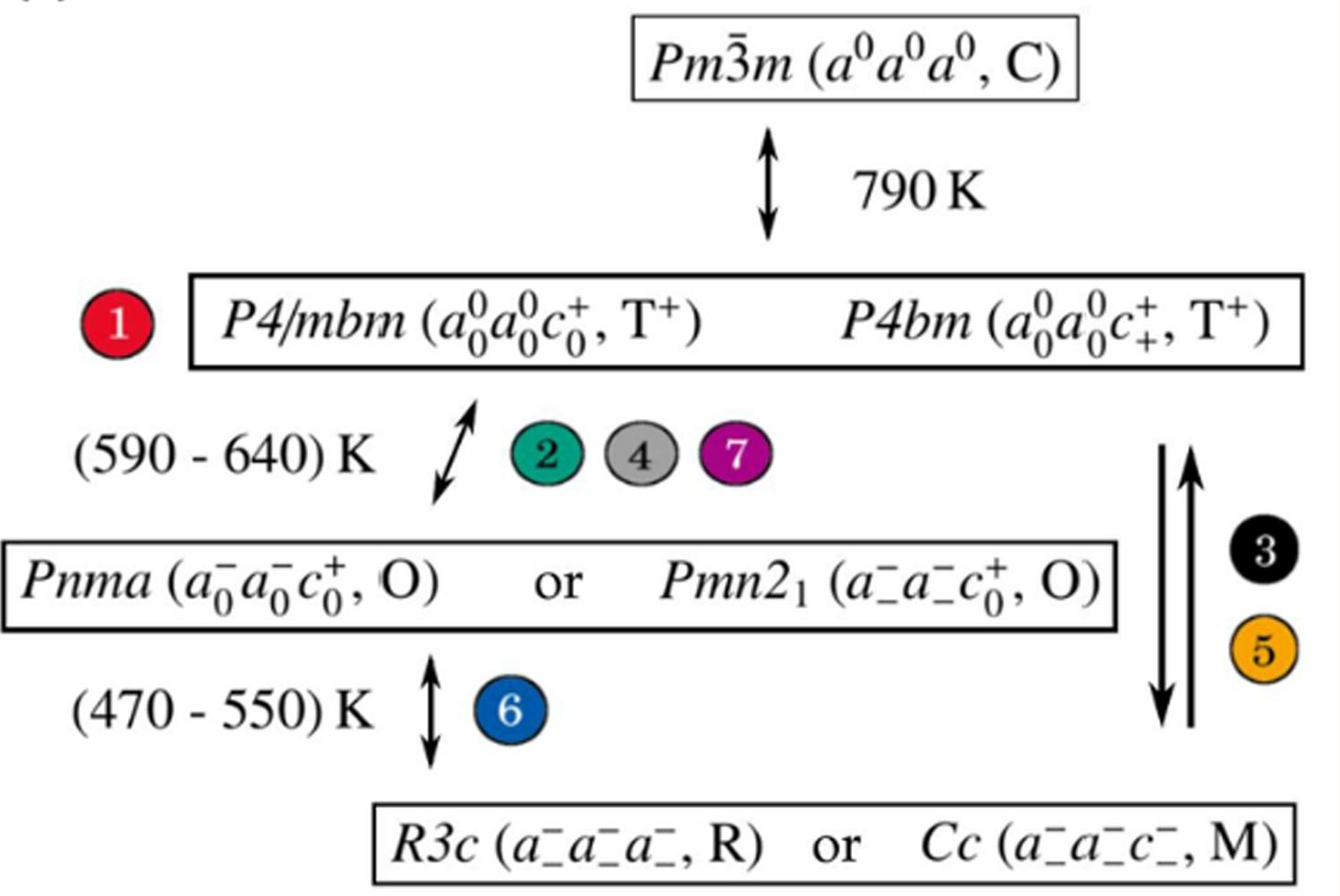}

}

\caption{Schematic diagram showing (a) AFE relaxor beahavior in temperature
range (200-230\textcelsius ). Reproduced with permission from Ref. \citep{Dorcet2008a}. Copyright 2008 American Chemical Society. and (b) domain switching
during entire phase change process of BNT. Reproduced with peermission from Ref. \citep{Meyer2018}. Copyright 2018 The American Ceramic Society. \label{fig:Sch} }

\end{figure}

The presence of polar nanoregions (PNRs) emerging from the A-site
disorder of Bi$^{3+}$ and Na$^{+}$ is ascribed to the relaxor action
of BNT. Notably, the dipolar freezing temperature $T_{f}$ of BNT
is 190$\lyxmathsym{\textcelsius}$ (near the $T_{d}$) \citep{Kreisel2002},
indicating that at room temperature, BNT exhibits a nonergodic relaxor
(\textquotedbl NR\textquotedbl ) state and behaves as a normal ferroelectric
with a \textquotedbl square\textquotedbl{} polarization-electric
field (P-E) loop, high remnant polarization $P_{r}$, and definite
macro-piezoelectricity $d_{33}$. BNT is in an ergodic relaxor (\textquotedbl ER\textquotedbl )
state above $T_{f}$ which also exhibits AFE relaxor behavior, and
an electric field cannot maintain long-range and stable ferroelectric
order. The AFE relaxor behavior in the ER phase for the temperature
range of 200-230$\lyxmathsym{\textcelsius}$ is shown in Fig. \ref{fig:Sch}(a).
The AFE relaxor behavior is caused by torsional strain resulting from
the reorientation of polar moments and octahedral tilting due to an
increase in temperature. This leads to the formation of a Pnma sheet
and a decrease in spontaneous polarization, which is more pronounced
at a temperature of 200$\lyxmathsym{\textcelsius}$. The value of
$T_{f}$ (or $T_{d}$) can be adjusted by composition structure design,
and the room-temperature relaxor ferroelectric (FE) phase of BNT-based
materials will thus be manipulated to be \textquotedbl NR\textquotedbl{}
phase, \textquotedbl NR\textquotedbl /\textquotedbl ER\textquotedbl{}
phase boundary, or \textquotedbl ER\textquotedbl{} phase, according
to the diverse functional qualities. The optimal setting for piezoelectric
applications is to have a high $d_{33}$ and $T_{d}$. As with the
PZT system, constructing an R/T multi-phase boundary (MPB) in BNT-based
compounds, such as the conventional BNT-BT system first reported by
Takenaka et al. \citep{Takenaka1991}, is an effective technique to
increase the $d_{33}$. Several explanations for this piezoelectric
enhancement have been offered, including two-phase coexistence \citep{Takenaka1991},
a decrease in lattice distortion as one approaches the phase border
\citep{Sung2010}, and the occurrence of an intermediate monoclinic
state \citep{Maurya2013}. However, due to the difficulty of polarization,
the $d_{33}$ of BNT-based systems is not very high $(d_{33}\sim300pc/N)$
to date, and there is an inverse relationship between $d_{33}$ and
$T_{d}$. Many techniques have been used to improve the $d_{33}$
and $T_{d}$ of BNT-based systems, including texture design \citep{Tam2008,Maurya2012,Bai2018,Maurya2015},
grain size control \citep{Khatua2019}, metal oxide addition \citep{Yin2018,Zhang2016,Zhang2015},
and quenching treatment \citep{Zhang2019,Yin2020}. The improved piezoelectric
response enable us to use BNT-based composites in high temperature
sensor, and haptic device applications.

BNT-based ceramics exhibit a border between the \textquotedbl NR\textquotedbl{}
(ferroelectric) and \textquotedbl ER\textquotedbl{} (nonpolar) phases
at temperatures near $T_{d}$, which encourages the formation of enormous
field-induced stresses due to a reversible nonpolar-ferroelectric
phase transition and domain switching \citep{Zhang2007,Hiruma2008,Aksel2012,Aksel2012a}.
BNT-based ceramics revealed a pinched P-E loop and a sprout-shaped
field-induced strain (S-E) curve, as well as a big maximum strain
$S_{max}$, a low residual strain $S_{rem}$, and a negative strain
$S_{neg}$. By lowering the $T_{d}$ below room temperature, an \textquotedbl ER\textquotedbl{}
state containing PNRs can be created in BNT-based ceramics, which
typically have a narrow P-E loop \citep{Malik2014,Liu2016}. To optimize
the recoverable energy density $J_{reco}$ and energy efficiency $\eta$,
dielectric materials should have a slim P-E loop with a large maximum
polarization $P_{max}$, low remenant polarization $P_{r}$, high
breakdown strength $E_{B}$, and low hysteresis \citep{Hao2013}.
All of these characteristics provides the adapatibility to BNT to
use as high temperature actuator applications during this regime.

BNT's phase transition procedure is more complex beyond the 320\textcelsius{}
\citep{S.B.Vakhrushev1985}. At around 320$\lyxmathsym{\textcelsius}$
(593.15K), BNT changes to a ferroelectric phase with tetragonal symmetry
P4bm. The P4bm symmetry exhibits in-phase a$^{0}$a$^{0}$c$^{+}$ octahedral
tilting (see Fig. \ref{fig:Sch}(b)), which is paired with anti-parallel
displacement of the Na$^{+}$/Bi$^{3+}$ and Ti$^{4+}$ cations along
[001]P$_{C}$, with only 0.2 percent distortion from the cubic structure
\citep{Jones2002,Jones2000}. As the temperature rises to around 520
$\lyxmathsym{\textcelsius}$, the crystal structure changes to cubic
Pm3m. One perspective is that the phase transition occurs about 520
\textcelsius{} between the cubic paraelectric and tetragonal ferroelastic
phases \citep{Geday2000}, while another view is that the phase transition
occurs between the cubic paraelectric and tetragonal super-paraelectric
phases \citep{Tu1994}. In this regime (320-520\textcelsius ), the
BNT posseses non-negligible hysteresis, and non-negligible $P_{r}$
\citep{Li2019,Qi2019} which provide a possibility to use this ferroelectric/super-paraelectric
phase for sensor and haptic applications at these high temperatures.
The BNT shows paraelectric phase beyond 520\textcelsius , therefore
negligible $P_{r}$ at this stage limits its use for sensor and haptic
applications.

Our breif literature review presented above above illustrates that
the thermo-electromechanical behavior of BNT is influenced by phase
and domain transitions at various temperatures. The electromechanical
response is not only dependent on temperature, but also on the phase
structure of BNT, which is crucial. Hence, this article presents a
temperature-dependent phase-field physics-based computational model
that aims to forecast the thermo-electromechanical behavior of BNT
across various temperature ranges. The validation of the numerical
phase-field thermo-electromechanical model will include replicating
the experimentally confirmed polarization potential values of BNT
material at various temperatures. The developed numerical model has
the ability to accurately anticipate the intricate phase transition
and domain switching behavior of BNT material. Consequently, it can
be used to forecast the precise thermo-electromechanical response,
enabling accurate determination of the appropriate application for
certain temperature regimes. The rest of the paper is stuctured as
follows: In section 2, we have provided the detailed constitutive
relations, governing equations, boundary conditions along with geometric
description and material properties. Section 3 provides the results
and discussions along with experimental validation of the developed
numerical model. Finally, section 4 can assist the conclusions and
possible future directions of the present study. Among other advantages,
this research will also aid in altering the domain structure by including
additional material based on the specific temperature range necessary
for a particular application. 

\section{Material and method}

The transient thermo-electromechanical behaviour with complex phase
change and domain switching for a two-dimensional piezoelectric composite
architecture comprised of microscale piezoelectric inclusions (BNT)
is investigated. The sub-sections that follow will go over the composite
architecture, the coupled thermo-piezoelectric model that was used
to study the composite's behaviour, the materials models that govern
the dielectric and mechanical properties of the composite, and the
boundary conditions that were used to compute specific effective electro-elastic
coefficients of interest.

\subsection{Geometrical description of piezoelectric composite and inclusions}

\begin{figure}
\subfloat[]{\includegraphics[scale=0.32]{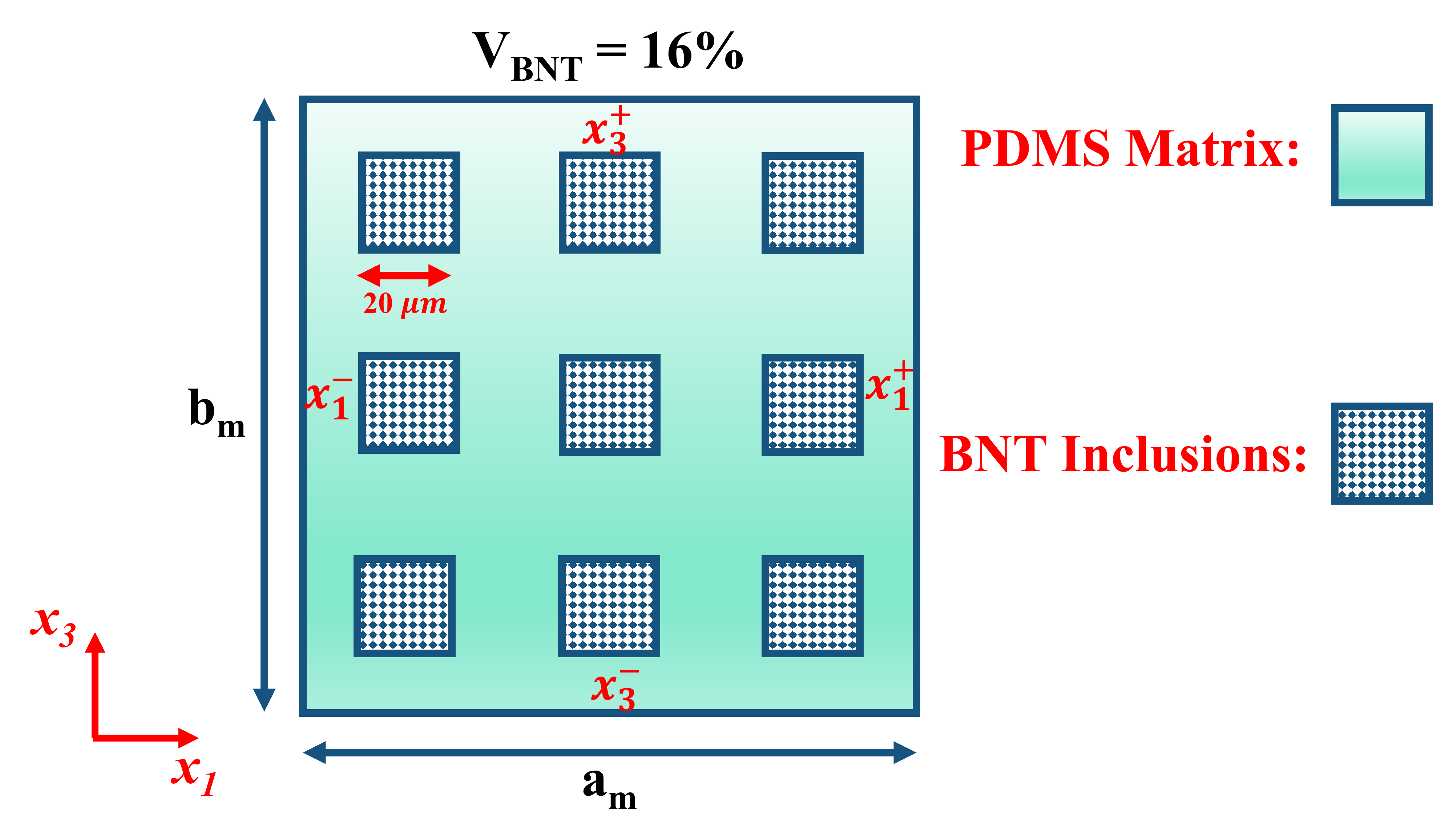}

}\subfloat[]{\includegraphics[scale=0.4]{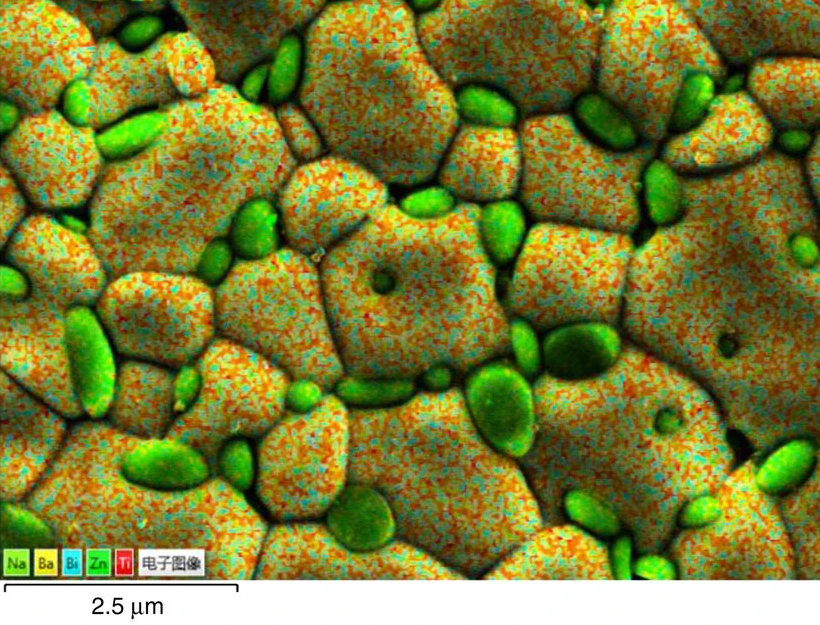}

}

\caption{(a) Piezocomposites with PDMS matrix and BNT inclusions, and (b) microstructure
images of BNT-6BT composites. Reproduced with permission from Ref. \citep{Zhang2015}. Copyright 2015 Nature Communications. \label{fig:Fig1}}

\end{figure}

Figure \ref{fig:Fig1} depicts the composite being represented as
a two-dimensional Representative Volume Element (RVE) in the $x_{1}-x_{3}$
plane.  The validity of the RVE assumptions is ensured by the existence of a homogenized matrix composed of the polymer and piezoelectric inclusions subject to periodic fields. This holds true for quasi-static loads, provided that the shortest characteristic length of the computational domain is significantly smaller than the wavelength associated with the frequency of the external mechanical stress \citep{Blanco2016,Krishnaswamy2021}. The composite consists of two components: (i) a rectangular
matrix with dimensions $a_{m}$ and $b_{m}$, and (ii) square-shaped
inclusions of microscale polycrystalline lead-free BNT. The square-shaped
piezoelectric inclusions have a side of approximately 20 $\mu m$,
whereas the sides ($a_{m}$ and $b_{m}$) of the shown RVE geometry
measure 150 $\mu m$ in length. The geometric representation is influenced
by our previous research  \citep{Krishnaswamy2021, Akshayveer2024a},
where we examined the geometry of matrices with identical sizes and a comparable
quantity of BaTiO$_{3}$ material inclusions \citep{Krishnaswamy2021} and BNT inclusions \citep{Akshayveer2024a}. Nevertheless, although
they employed inclusions of varying sizes \citep{Krishnaswamy2021}, in what follows, we utilized
BNT inclusions with a squared shape \citep{Akshayveer2024a}. Consequently, the volume fraction
differs slightly despite having the same number of inclusions. The
volume fraction of BNT inclusions is contingent upon the quantity
of inclusions. The piezoelectric inclusions have a microscopic length
scale, while the grain size has a nano-range length scale (see Fig.
\ref{fig:Fig1}(b)). Hence, it is crucial to note that the size of
the inclusions should exceed the optimal grain size of the composite
in order to achieve improved piezoelectric properties in the poly-crystal
of BNT.   

\subsection{Mathematical model}

This work focuses on analyzing the composite geometry discussed in
the previous section. We present the thermo-electromechanical model
employed to analyze the composite design discussed in section 2.1
and visually shown in Fig. \ref{fig:Fig1}. The phase transitions
and ferroelectric domain transitions are considered in the model by
incluiding the Landau-Ginzburg-Devonshire contribution (in \citep{Ahluwalia,Morozovska,Wang, Akshayveer2024})
to (the first two terms of the Eq. (\ref{eq:free_energy})) the Helmholtz
free energy ($\phi(\pmb{\overrightarrow{E}},\pmb{\varepsilon}(\pmb{\overrightarrow{u}}), \nabla\pmb{\varepsilon}, \pmb{\overrightarrow{p}},\pmb{\nabla\overrightarrow{p}},\theta)$)
of the system \citep{Borrelli2019}, which is described as:
\begin{eqnarray}
\phi(\pmb{\overrightarrow{E}},\pmb{\varepsilon}(\pmb{\overrightarrow{u}}),\nabla\pmb{\varepsilon}, \pmb{\overrightarrow{p}},\pmb{\nabla\overrightarrow{p}},\theta) & = & \frac{1}{2}\lambda\left|\pmb{\nabla\overrightarrow{p}}\right|^{2}+W^{3D}(\theta,\pmb{\overrightarrow{p}})+\frac{1}{2}\pmb{\varepsilon^{el}:C\varepsilon^{el}}-(\theta-\theta_{R})\pmb{\beta:\varepsilon}(\pmb{\overrightarrow{u}})-\frac{1}{2}\pmb{\overrightarrow{E}}^{T}\pmb{\epsilon{\overrightarrow{E}}}\nonumber \\
 & - &\pmb{\mu\overrightarrow{E}\nabla\varepsilon^{el}}-\pmb{e\overrightarrow{E}\varepsilon^{el}}-\pmb{\overrightarrow{p}\overrightarrow{E}}-\pmb{\eta}(\theta-\theta_{R})\pmb{\overrightarrow{E}}.\label{eq:free_energy}
\end{eqnarray}
The first term of the Eq. (\ref{eq:free_energy}) is the Landau gradient
energy term due to spatial gradient of polarization vector $\pmb{\overrightarrow{p}}$ ($[p_{1} \quad p_{3}]^{T}$) through the two-dimensional domain,
second term is the Landau-Ginzburg-Devonshire function of free energy
(see Eq. (\ref{eq:polynomial}) below), and the remaining terms
are elastic deformation energy, thermal deformation energy, dielectric
energy, flexoelectric energy, electromechanical energy, electric polarization
energy, and thermoelectric energy respectively. The variables
$\pmb{C}$, $\pmb{\epsilon}$, $\pmb{e}$, $\pmb{\mu}$, $\pmb{\beta}$,
and $\pmb{\eta}$ represent the elastic constants, dielectric
constants, piezoelectric constants, flexoelectric constants, thermal
expansion coefficient, and thermoelectric constants respectively. Moreover, $\pmb{\overrightarrow{E}}$ is the
electric field intensity vector,
$\pmb{\varepsilon(\overrightarrow{u})}$ is total mechanical strain
tensor, $\pmb{\overrightarrow{u}}$ is the mechanical displacement vector.
$\theta$ is the temperature, and $\theta_{R}$ is the reference temperature.
The overall mechanical strain $\pmb{\varepsilon(\overrightarrow{u})}$ consists of two
components: the elastic strain tensor $\pmb{\varepsilon^{el}}$, which
is directly proportional to the stress $\pmb{\sigma}$, and the transformation
strain tensor $\pmb{\varepsilon^{t}(\overrightarrow{p})}$, which is
connected to phase change and ferroelectric transition.
\begin{eqnarray}
	\pmb{\varepsilon}(\pmb{\overrightarrow{u}}) & = & \frac{1}{2}(\pmb{\nabla\overrightarrow{u}+\nabla \overrightarrow{u}}^T)=\pmb{\varepsilon^{el}}+\pmb{\varepsilon^{t}}(\pmb{\overrightarrow{p}}),\label{eq:total_strain}
\end{eqnarray}
where $\pmb{\varepsilon^{t}(\overrightarrow{p})}$ is a result of the distortion
of the crystal unit cell caused by the ferroelectric transition. The
tensor reflects a deformation that occurs in the direction, which
is aligned with the ferroelectric polarization $\pmb{\overrightarrow{p}}$. This deformation
is directly proportional, through a material constant $\gamma$, to
the modulus of $\pmb{\overrightarrow{p}}$ \citep{Kamlah1999}
and shown as:
\begin{eqnarray}
\pmb{\varepsilon^{t}}(\pmb{\overrightarrow{p}}) & = & \gamma\left|\pmb{\overrightarrow{p}}\right|(\pmb{\overrightarrow{n}}\otimes\pmb{\overrightarrow{n}}-\frac{1}{d}\pmb{I}),\pmb{\overrightarrow{n}}\coloneqq\frac{\pmb{\overrightarrow{p}}}{\left|\pmb{\overrightarrow{p}}\right|}.\label{eq:transformation_strain}
\end{eqnarray}
where $d=3$ for three-dimensional deformations \citep{Kamlah2001}. In the case of plane strain, where we set $n_2 = 0$ and $\varepsilon_{22}=0$, the factor $d$ should be taken as $d=2$. More precisely, the transformation strain involves an elongation that is proportional to $\gamma |\pmb{\overrightarrow{p}}|$ in the direction of the polarization, $\mathbf{\pmb{\overrightarrow{n}}}$. The direction perpendicular to $\mathbf{\pmb{\overrightarrow{n}}}$ within the plane experiences a decrease in length, while the out-of-plane direction remains unchanged. This adjustment for $d$ ensures there is no volume change in plane strain scenarios. It is crucial to note that the transformation strain exhibits
symmetry in relation to the unit vector $\pmb{\overrightarrow{n}}$, indicating
that the same deformation is linked to different polarization states.
Therefore, domains that have a combination of positive and negative
polarization states $(\pm\pmb{\overrightarrow{n}})$ can have a polarization
that becomes zero, while still having a non-zero transformation strain.
Since $\pmb{\overrightarrow{n}}\otimes\pmb{\overrightarrow{n}}$ is the tensor
multiplication of unit vector to its transpose vector; therefore,
the transformation strain is symmetric with regard to $\pmb{\overrightarrow{n}}$,
resulting in opposing polarization states for the same deformation.
As a result, the strain displays a shape like a butterfly and a hysteresis
curve for the polarization vector when an oscillating electric field
is applied \citep{Landis2002,McMeeking2002}.

The Landau-Ginzburg-Devonshire free energy function $W^{3D}(\theta,\pmb{\overrightarrow{p}})$
characterizes the temperature-induced phase change, as well as other
thermal characteristics of the material, such as specific heat. As
per Landau's recommendations, the Landau-Ginzburg-Devonshire free
energy takes the shape of a polynomial function in the order parameter
$\pmb{\overrightarrow{p}}$, meeting specific symmetry criteria based on the
material's symmetry. For a three-dimensional anisotropic crystal exhibiting, the following
Landau-Ginzburg-Devonshire free energy function represents a polynomial \citep{Voelker2011,Indergand2020,Indergand2021} is defined as:
\begin{eqnarray}
W^{3D}(\theta,\pmb{\overrightarrow{p}}) & = &\alpha_{1}\frac{\theta_{c}-\theta}{\theta_{c}}(p_{1}^{2}+p_{2}^{2}+p_{3}^{2})+\alpha_{11}(p_{1}^{4}+p_{2}^{4}+p_{3}^{4})+\alpha_{12}\frac{\theta_{c}-\theta}{\theta_{c}}(p_{1}^{2}p_{2}^{2}+p_{2}^{2}p_{3}^{2}+p_{3}^{2}p_{1}^{2})\nonumber \\
 & + &\alpha_{111}(p_{1}^{6}+p_{2}^{6}+p_{6}^{4})+ \alpha_{112}[p_{1}^{4}(p_{2}^{2}+p_{3}^{2})+p_{2}^{4}(p_{3}^{2}+p_{1}^{2})+p_{3}^{4}(p_{1}^{2}+p_{2}^{2})]\nonumber\\&+&\alpha_{123}(p_{1}^{2}p_{2}^{2}p_{3}^{2}),\label{eq:polynomial}
\end{eqnarray}
with $\alpha_{1}$, $\alpha_{11}$, $\alpha_{12}$, $\alpha_{111}$, $\alpha_{112}$ and $\alpha_{123}$ are material characteristics and $\theta_{c}$ represents the Curie temperature.
Devonshire \citep{Devonshire1949}
only considers the term $\alpha_{111}(p_{1}^{6}+p_{2}^{6}+p_{3}^{6})$ in
the sixth-order expansion because of the negligible impact of other
sixth-order terms. The polynomial $W^{3D}(\theta,\pmb{\overrightarrow{p}})$ can be written as:
\begin{eqnarray}
W^{3D}(\theta,\pmb{\overrightarrow{p}}) & = &\alpha_{1}\frac{\theta_{c}-\theta}{\theta_{c}}(p_{1}^{2}+p_{2}^{2}+p_{3}^{2})+\alpha_{11}(p_{1}^{4}+p_{2}^{4}+p_{3}^{4})+\alpha_{12}\frac{\theta_{c}-\theta}{\theta_{c}}(p_{1}^{2}p_{2}^{2}+p_{2}^{2}p_{3}^{2}+p_{3}^{2}p_{1}^{2})\nonumber \\
 & + &\alpha_{111}(p_{1}^{6}+p_{2}^{6}+p_{3}^{6}).\label{eq:polynomial2}
\end{eqnarray}
The minimization of Helmholtz free eneergy function $\phi$ with respect to  $\pmb{\overrightarrow{p}}$ will also leads to minimization of $W^{3D}(\theta,\pmb{\overrightarrow{p}})$, which has a global minimum at $\pmb{\overrightarrow{p}}=0$ for every $\theta\geq\theta_{0}$ in absence of electric field. Where, $\theta_{0}$ represents the transition temperature (close to
to Curie temperature $\theta_{c}$). 

Moreover, it is postulated that the polarization component along the translation axis is inconsequential ($p_{2} = 0$). In these instances, a perovskite ferroelectric that experiences a transition from a cubic to a tetragonal phase can be considered as an additional two-dimensional ferroelectric with a transition from a square to a rectangle phase and similar 2D transitions in other phase transformations \citep{Ahluwalia,Morozovska,Wang}. For the present 2-dimensional $x_{1}-x_{3}$ computational domain, the polynomial $W^{2D}(\theta,\pmb{\overrightarrow{p}})$ is written as:
\begin{eqnarray}
W^{2D}(\theta,\pmb{\overrightarrow{p}})  = \alpha_{1}\frac{\theta_{c}-\theta}{\theta_{c}}(p_{1}^{2}+p_{3}^{2})+\alpha_{11}(p_{1}^{4}+p_{3}^{4})+\alpha_{12}\frac{\theta_{c}-\theta}{\theta_{c}}p_{3}^{2}p_{1}^{2}+\alpha_{111}(p_{1}^{6}+p_{3}^{6}).\label{eq:polynomial2D}
\end{eqnarray}
The calculations regarding Landau-Ginzburg-Devonshire free energy function have been presented in \ref{A}, \ref{B}, and \ref{C}.

As mentioned, the order parameter used to describe the phase transition is the ferroelectric
polarization vector $\pmb{\overrightarrow{p}}$. The transient Landau-Ginzburg-Devonshire
equation for a vector order parameter $\pmb{\overrightarrow{p}}$ is
a semilinear parabolic equation, often known as a reaction-diffusion
system \citep{Kamlah2001,Mueller2008} and described as:
\begin{eqnarray}
\tau\dot{\pmb{\overrightarrow{p}}} & = & \nabla\cdotp(\lambda\nabla\pmb{\overrightarrow{p}})-W^{2D}_{\pmb{\overrightarrow{p}}}(\theta,\pmb{\overrightarrow{p}}),\label{eq:parabolic}
\end{eqnarray}
where $\tau$ and $\lambda$ are the material properties. The
first term of the right hand side of Eq. (\ref{eq:parabolic}) indicates
the non-uniform polarization distribution for nano-scale ferroelectric
distribution, which is neglected for the the present investigation,
but it can be considered for future studies. To determine the implications
of the second law of thermodynamics, one can express the dissipation
inequality for the Helmholtz free energy function $(\phi)$ by combining
the energy balance and entropy inequality:
\begin{eqnarray}
 & (\phi_{\theta}+S)\dot{\theta}+(\phi_{\pmb{\overrightarrow{p}}}-W^{2D}_{\pmb{\overrightarrow{p}}})-\mathcal{D}_{\pmb{\overrightarrow{p}}})\cdotp\dot{\pmb{\overrightarrow{p}}}+(\phi_{\pmb{\varepsilon}}-\pmb{\sigma})\cdotp\dot{\pmb{\varepsilon}}\nonumber \\
 & +(\phi_{\pmb{\nabla\varepsilon}}-\pmb{\hat{\sigma}})\cdotp\nabla\dot{\pmb{\varepsilon}}+(\phi_{\pmb{\overrightarrow{E}}}+\pmb{\overrightarrow{D}})\cdotp\dot{\pmb{\overrightarrow{E}}}+\frac{\pmb{\overrightarrow{q}}}{\theta}\cdotp\nabla\theta\leq0,\label{eq:inequality}
\end{eqnarray}
where $\mathcal{D}_{\pmb{\overrightarrow{p}}}$ is the derivative of dissipation potential
$\mathcal{D}(\dot{\pmb{\overrightarrow{p}}}$) and defined as $\mathcal{D}_{\dot{\pmb{\overrightarrow{p}}}}=\tau\dot{\pmb{\overrightarrow{p}}}$ which
is immediately seen to be sufficient to satisfy the dissipation inequality
in Eq. (\ref{eq:Rest_eq}), and $\pmb{\overrightarrow{q}}$ is the heat flux vector, and
$S$ is the entropy. The heat flux ($\pmb{\overrightarrow{q}}$) and electric field vector ($\pmb{\overrightarrow{E}}$) can be defined as $\pmb{\overrightarrow{q}}=-\pmb{k}\nabla\theta$, and $\pmb{\overrightarrow{E}}=-\nabla V$ respectively. Moreover, the following constitutive equations
can be derived from the Eq. (\ref{eq:free_energy}):
\begin{eqnarray}
\phi_{\pmb{\varepsilon}}=\pmb{\sigma}=\pmb{C}(\pmb{\varepsilon}(\pmb{\overrightarrow{u}})-\pmb{\varepsilon^{t}}(\pmb{\overrightarrow{p}}))-\pmb{e}\pmb{\overrightarrow{E}}-\pmb{\beta}(\theta-\theta_{R}),\label{eq:Stress}
\end{eqnarray}
\begin{eqnarray}
\phi_{\pmb{\nabla\varepsilon}} & = & \pmb{\hat{\sigma}}=\pmb{\mu}\pmb{\overrightarrow{E}},\label{eq:flexo}
\end{eqnarray}
\begin{eqnarray}
-\phi_{\pmb{\overrightarrow{E}}} =  \pmb{\overrightarrow{D}}&=&\pmb{\epsilon}\pmb{\overrightarrow{E}}+\pmb{e}(\pmb{\varepsilon}(\pmb{\overrightarrow{u}})-\pmb{\varepsilon^{t}}(\pmb{\overrightarrow{p}}))+\pmb{\eta}(\theta-\theta_{R})\nonumber \\
&+&\pmb{\mu}(\pmb{\nabla\varepsilon}(\pmb{\overrightarrow{u}})-\pmb{\nabla\varepsilon^{t}}(\pmb{\overrightarrow{p}}))+\pmb{\overrightarrow{p}},\label{eq:ED}
\end{eqnarray}
\begin{eqnarray}
-\phi_{\theta} & = & S=\pmb{\beta}\pmb{\varepsilon}(\pmb{\overrightarrow{u}})-\pmb{\eta}\pmb{\overrightarrow{E}}+W_{\theta}(\theta,\pmb{\overrightarrow{p}}),\label{eq:entropy}
\end{eqnarray}
\begin{eqnarray}
 &  & \phi_{\pmb{\overrightarrow{p}}}=W^{2D}_{\theta,\pmb{\overrightarrow{p}}}(\pmb{\overrightarrow{p}}),\mathcal{D}_{\dot{\pmb{\overrightarrow{p}}}}\cdotp\dot{\pmb{\overrightarrow{p}}}\geq0,\label{eq:Rest_eq}
\end{eqnarray}
where the condition $\mathcal{D}_{\dot{\pmb{\overrightarrow{p}}}}\cdotp\dot{\pmb{\overrightarrow{p}}}\geq0$
indicates that $\tau>0$. Now, we rephrase the system of balancing
equations by using the above constitutive laws. The variations in the micro-scale inertia and body forces with respect to their RVE volume averages have the potential to influence the micro-scale equilibrium problem and the homogenized stress that is produced as a consequence. When it comes to mechanical relevance, the volume average itself is only meaningful at the macroscale. The microstructure is significantly smaller than the external load's frequency, as a quasistatic load is applied here. Furthermore, the dynamics of phase transformation do not include the micro-inertia \citep{deSouzaNeto2015}. Therefore, the effect of micro-intertia terms and body forces are neglected for this study. The temperature of all sides of RVE are considered to be the same with no additional heat flux and internal heat generation. The governing equations
for the 2D composite system ($x_{1}-x_{3}$ plane) that describes
the model can be characterized as follows: 
\begin{eqnarray}
\nabla\cdotp(\pmb{\sigma}-\pmb{\hat{\sigma}})& = &  0.,\label{eq:Mechanical_ge}
\end{eqnarray}
\begin{eqnarray}
\tau\dot{\pmb{\overrightarrow{p}}} & = & -W^{2D}_{\theta,\pmb{\overrightarrow{p}}}(\pmb{\overrightarrow{p}}),\label{eq:parabolic-1}
\end{eqnarray}
\begin{eqnarray}
\nabla\cdotp(\pmb{k}\nabla\theta) = 0,\label{eq:heat2}
\end{eqnarray}
\begin{eqnarray}
\nabla\cdotp \pmb{\overrightarrow{D}} & = & 0,\label{eq:Electric_ge}
\end{eqnarray}

\subsection{Boundary and initial conditions}

The composite geometry as schematically represented in Fig. \ref{fig:Fig1} is a two-dimensional approximation in the $x_{1}-x_{3}$
plane. The RVE approximation is used for this model as there is a homogenized matrix of the polymer and piezoelectric inclusions. This assumption is valid for quasi-static loads, where we assume that the smallest characteristic length of the computational domain is significantly smaller than the wavelength corresponding to the frequency of the external mechanical load. However, the periodic boundary conditions can lead to a better estimation of the effective properties of the material. Moreover, the boundary conditions applied in this study can lead to similar effective properties; hence, the RVE approximation is valid for these types of boundary conditions \citep{Krishnaswamy2020a,Blanco2016}. Thus, the governing equations that regulate the thermo-electromechanical
characteristics of the composite material are formulated based on
the 2-D composite geometry being studied, together with the constitutive
laws. As a first step, we will examine the P-E characteristics curve and the S-E characteristics curve for the composite material in order to assess its piezoelectric and ferroelectric phase-transition behaviour. The determination of these curves requires the application periodic boundary conditions, which are explained in Eq. (\ref{BC}) correspondingly for all variables.
\begin{eqnarray}
u_{1}(x_{1}^{+},t)-u_{1}(x_{1}^{-},t)=\varepsilon_{0}\cdot(x_{1}^{+}-x_{1}^{-}),~u_{1}(x_{3}^{+},t)=u_{1}(x_{3}^{-},t)=0\nonumber\\
u_{3}(x_{3}^{+},t)=u_{3}(x_{3}^{-},t)=u_{3}(x_{1}^{+},t)=u_{3}(x_{1}^{-},t)=0,\nonumber\\
V(x_{3}^{+},t)=V(x_{3}^{-},t)=E_{0}\cdot(x_{3}^{+}-x_{3}^{-}),~V(x_{1}^{+},t)=V(x_{1}^{-},t)=0,\nonumber\\
p_{1}(x_{1}^{+},t)=p_{1}(x_{1}^{-},t),\nonumber\\
p_{3}(x_{3}^{+},t)=p_{3}(x_{3}^{-},t),\nonumber\\
\theta(x_{3}^{+},t)=\theta(x_{3}^{-},t)=\theta(x_{1}^{+},t)=\theta(x_{1}^{-},t)=\theta_{S},\label{BC}
\end{eqnarray}
where $\varepsilon_{0}$ and $E_{0}$ are average values of externally applied mechanical strain and electric field, whose values are $1.0\cdot e^{-6}$ and $3.2\cdot e^{7}~ N/C$ respectively \citep{Indergand2021, Indergand2020}. In order for homogenization to be effective, we make the assumption that there is a clear distinction between different scales and we propose that the forces exerted by the body and inertia may be disregarded. This is because we are primarily concerned with the quasi-static behaviour of the material. As a result, the mechanical and electrical issues related to the Representative Volume Element (RVE) are tackled using quasi-static methods. 
The governing equation for polarization vector (Eq.(\ref{eq:parabolic-1})) that depends on time is solved by assuming periodic boundary conditions, which do not enforce an average but instead permit the polarization field to develop without constraints, save for periodicity on the surfaces of the RVE. 

The value of the reference temperature $\theta_{R}$ is assumed to
be the ambient temperature which is $27\text{\textcelsius},$ and the
thermal boundary conditions at the all walls is above reference temperature
and designated as $\theta_{S}$ which can vary up to 320\textcelsius.  Apart from the boundary conditions, the initial
conditions also play a vital role in the transient phase-field thermo-electromechanical
modelling of BNT-based piezoelectric composite. The initial conditions
are assumed as:
\begin{eqnarray}
u_{1}(t=0)=0, u_{3}(t=0)=0, V(t=0)=-E_{0}\cdot x_{3}, ~\theta(t=0)=\theta_{R}, p_{1}=0, p_{3}=-p_{s}.\label{eq:Initial_cond}
\end{eqnarray}

\subsection{Effective material coefficients}

The RVE approximation is employed in these model due to the presence of a homogenized matrix
consisting of both the polymer and piezoelectric inclusions. To transit from the microscale to the macroscale, the effective response of a RVE is computed, using the conventional approach of classical first-order homogenization. To characterize a sample with a microstructure that is nearly statistically homogenous, we define an effective property as the volume average over the RVE. The volume average of quantity $A$ is denoted as $\left\langle A\right\rangle $
in the following calculation. It is computed as \citep{Schroeder2009,Miehe2002}:
\begin{eqnarray}
\left\langle A\right\rangle  & = & \frac{1}{(a_{m}b_{m})}\int_{\Omega}Ad\Omega,\label{eq:RVE}
\end{eqnarray}
where $\Omega$  is the volume across which the integration is
performed, the complete volume of the RVE in this example. The effective mechanical properties for homogenized RVE computational domain can be calculated as:
\begin{eqnarray}
C_{ij,eff.}=\frac{\left\langle \sigma_{ii}\right\rangle }{\varepsilon_{jj}}, & \text{\text{\ensuremath{e_{ij,eff.}}=\ensuremath{\frac{\left\langle D_{i}\right\rangle }{\varepsilon_{jj}}}},}  & for~i,j =1,3.\label{eq:Effect}
\end{eqnarray}
In this model, the applied displacement boundary conditions on the RVE are computed under the assumption of infinitesimal strains, 
$\varepsilon_{jj}$. This hypothesis is consistent with the linearization of the material properties, $C_{ij,eff}$ and $e_{ij,eff}$, at zero deformation, ensuring that the computed effective properties remain within the linear elastic regime. In Eq. (\ref{eq:Effect}) the effective piezoelectric coefficient ($e_{ij,eff.}$) is caluculated using stress form equations, while strain form piezoelectric coefficient ($d_{ij,eff.}$) is very important as most of the experimental literature exhibit $d_{ij,eff.}$ only. The piezoelectric coefficients in strain form have been transformed into stress form using the formula $e_{ijk}=C_{jklm}d_{ilm}$ \citep{Singh2020}. Hence, the effective piezoelectric coefficient ($d_{ij,eff.}$) can be calculated as:
\begin{eqnarray}
d_{ij,eff.}=\frac{e_{ij,eff.}}{C_{ij,eff.}}.
\end{eqnarray}

\subsection{Material properties}

We selected polydimethylsiloxane (PDMS) as the soft polymer matrix
with micro-scale BNT inclusions for our computational experiment.
The literature has been paying more and more attention to experimental
efforts to create piezoelectric composites with soft matrices through
3D printing and other cutting-edge technologies \citep{Kim2014}.
These soft matrices provide two challenges, however. Initially, the
applied mechanical stimuli from the BNT are screened by their smooth
elastic characteristics, which leads to a relatively minimal production
of electric flux. Second, polymeric materials often have poor
dielectric strength, which hinders the electric flux from exiting
the piezoelectric inclusions. Additionally, the PDMS matrix can be
strengthened by incorporating a small amount of carbon nanotubes,
which improves its elastic strength and dielectric permittivity \citep{Fan2018,Pinming2020}.
It can be readily 3D-printed. These characteristics make it a suitable
choice for the polymer matrix in this study. The future focus of this
study will be to enhance the thermo-electromechanical performance
of the BNT-PDMS composite by adding carbon nanotubes, and the findings
of this study can serve as a reference for future comparative
studies. Our choice of micro-scale BNT piezoelectric inclusions allows
for grain sizes that maximize polycrystallinity and piezoelectric
response. The material properties of the PDMS matrix and BNT inclusions
at the specified reference temperature ($\theta_{R}$) are listed
in Table \ref{tab:Material-properties-required}. According to Hiruma
et al. \citep{Hiruma2009}, all material constants are expected to
change with temperature. There is not significant literature about
the values of the flexoelectric coefficients of BNT. Therefore, the
constant value of longitudinal and transverse flexoelectric coefficients
of order $10^{-6}$ and negligible shear flexoelectric coefficient
(range for BaTiO$_{3}$) is considered for this study \citep{JAGDISH2024}. Analysis of
the thermal stability and degradation of polymeric composites based
on PDMS is a crucial problem that should be examined at higher temperatures.
\begin{table}[!t]
\centering
	\caption{Material properties required for the simulations. \label{tab:Material-properties-required}}
\begin{tabular}{|c|c|c|}
\hline 
\textbf{\footnotesize{}Material Property} & \textbf{\footnotesize{}BNT }{\footnotesize{}\citep{Hiruma2009}}\textbf{\footnotesize{}:} & \textbf{\footnotesize{}PDMS \citep{Krishnaswamy2020}:}\tabularnewline
\hline 
\textbf{\footnotesize{}Elastic coefficients (GPa):} &  & \tabularnewline
{\footnotesize{}$c_{11}$ } & {\footnotesize{}153.9 } & {\footnotesize{}$\lambda_{m}+2\mu_{m}$}\tabularnewline
{\footnotesize{}$c_{13}$ } & {\footnotesize{}52.1 } & {\footnotesize{}$\lambda_{m}$}\tabularnewline
{\footnotesize{}$c_{33}$ } & {\footnotesize{}168.1} & {\footnotesize{}$\lambda_{m}+2\mu_{m}$}\tabularnewline
{\footnotesize{}$c_{44}$ } & {\footnotesize{}82.3} & {\footnotesize{}$\mu_{m}$}\tabularnewline
{\footnotesize{}Young\textquoteright s Modulus ($E_{m}$) } & {\footnotesize{}93.0 } & {\footnotesize{}0.002}\tabularnewline
{\footnotesize{}Poisson\textquoteright s Ratio ($\nu_{m}(-)$)} & {\footnotesize{}0.23 } & {\footnotesize{}0.499}\tabularnewline
\hline 
\textbf{\footnotesize{}Relative permittivity: } &  & \tabularnewline
{\footnotesize{}$\frac{\epsilon_{11}}{\epsilon_{0}}$} & {\footnotesize{}367 } & {\footnotesize{}2.72}\tabularnewline
{\footnotesize{}$\frac{\epsilon_{33}}{\epsilon_{0}}$} & {\footnotesize{}343 } & {\footnotesize{}2.72}\tabularnewline
\hline 
\textbf{\footnotesize{}Piezoelectric coefficients $(pC/N)$:} &  & \tabularnewline
{\footnotesize{}$d_{15}$ } & {\footnotesize{}87.3 } & {\footnotesize{}Non-piezoelectric}\tabularnewline
{\footnotesize{}$d_{31}$ } & {\footnotesize{}-15.0 } & \tabularnewline
{\footnotesize{}$d_{33}$ } & {\footnotesize{}72.9 } & \tabularnewline
\hline 
\textbf{\footnotesize{}Flexoelectric coefficients $(cm^{-1})$:} &  & \tabularnewline
{\footnotesize{}$\mu_{11}$, longitudinal } & {\footnotesize{}$10^{-6}$} \citep{Narvaez2015} & {\footnotesize{}$10^{-9}$ }\citep{Wen2019}\tabularnewline
{\footnotesize{}$\mu_{12}$, transverse } & {\footnotesize{}$10^{-6}$ }\citep{Narvaez2015} & {\footnotesize{}$10^{-9}$ }\citep{Wen2019}\tabularnewline
{\footnotesize{}$\mu_{44}$, shear} & {\footnotesize{}$0$ }\citep{Narvaez2015} & {\footnotesize{}0 }\citep{Wen2019}\tabularnewline
\hline 
\textbf{\footnotesize{}Thermo-elastic coefficients:} &  & \tabularnewline
{\footnotesize{}$\pmb{\beta}$} & {\footnotesize{}$2\times10^{-5}$.$K^{-1}$\citep{Suchanicz2010}} & {\footnotesize{}3.$2\times10^{-4}$$K^{-1}$}\tabularnewline
\hline 
\textbf{\footnotesize{}Thermoelectric coefficient: } &  & \tabularnewline
{\footnotesize{}$\pmb{\eta}$ } & {\footnotesize{}$27.3\times10^{-4}$$C/m^{2}K$ \citep{Liu2015}} & {\footnotesize{}Non-thermoelectric}\tabularnewline
\hline 
\textbf{\footnotesize{}Heat capacity coefficient:}{\footnotesize{} } &  & \tabularnewline
{\footnotesize{}$a_{T}$} & {\footnotesize{}500 $J/kgK$ \citep{Liu2015}} & {\footnotesize{}1460 $J/kgK$}\tabularnewline
\hline 
\textbf{\footnotesize{}Thermal conductivity:}{\footnotesize{} } &  & \tabularnewline
{\footnotesize{}$\pmb{k}$} & {\footnotesize{}1.48 $W/mK$ \citep{Liu2015}} & {\footnotesize{}0.27 $W/mK$}\tabularnewline
\hline 
\textbf{\footnotesize{}Spontaneous polarization:}{\footnotesize{}} &  & \tabularnewline
{\footnotesize{}$p_{s}$} & {\footnotesize{}38 $\mu C/cm^{2}$ \citep{JAFFE1958}} & {\footnotesize{}0.0 $\mu C/cm^{2}$}\tabularnewline
\hline 
\textbf{\footnotesize{}Landau coefficients:}{\footnotesize{}} &  & \tabularnewline
{\footnotesize{}$\alpha_{1}$} & {\footnotesize{}-11.1 $Nm^{2}/C^{-2}$ \citep{Wang2019}} & {\footnotesize{}0.0 $Nm^{2}/C^{-2}$}\tabularnewline
{\footnotesize{}$\alpha_{11}$} & {\footnotesize{}-69.3 $Nm^{6}/C^{-4}$ \citep{Wang2019}} & {\footnotesize{}0.0 $Nm^{6}/C^{-4}$}\tabularnewline
{\footnotesize{}$\alpha_{12}$} & {\footnotesize{}234.3 $Nm^{6}/C^{-4}$ \citep{Wang2019}} & {\footnotesize{}0.0 $Nm^{6}/C^{-4}$}\tabularnewline
{\footnotesize{}$\alpha_{111}$} & {\footnotesize{}803.8 $Nm^{10}/C^{-6}$ \citep{Wang2019}} & {\footnotesize{}0.0 $Nm^{10}/C^{-6}$}\tabularnewline
\hline 
\end{tabular}

\end{table}

All the above phenomenological relations are linked with the governing
equations and discretized for the present computational domain. The
computational area has been divided into a suitable number of diverse
triangular mesh components based on a mesh convergence analysis. Analysis
was conducted on the remnant polarization $P_{r}$ at various temperatures
for different number of grids. It was shown that there is little change
in $P_{r}$ after 38470 grids, suggesting that this is the ideal number
of grids. The grid layouts have mesh sizes ranging from 1.25$\mu m$
and 10$nm$, which are used in the current investigation. The discretized
equations are solved using the finite element technique for the computational
domain with specified boundary conditions for the BNT material.

\section{Results and discussions}

The complex multiphase properties of lead-free piezoelectric materials,
such as BNT, enhance their adaptability for applications in environmentally
conscious and sustainable technologies. BNT exhibits intricate ferroelectric
domain switching over different temperature ranges, leading to the
existence of several phases and coexistence in complex multiphase
states. The behaviour of BNT material exhibits diverse polarization
values and strain response at different temperatures, rendering it
appropriate for utilization as a sensor, actuator, and haptic device
in computer-human interactions across multiple temperature ranges.
Here, we provide a selection of research studies that examine the
impact of various thermal, electrical and mechanical conditions on different parameters such as $p_{3}$, $p_{1}$, $\varepsilon_{33}$, $\varepsilon_{11}$, and $V$, for BNT in order to assess
its potential as a versatile and environmentally friendly material.
Significant values of $p_{3}$, $p_{1}$ and $V$,  indicate the appropriateness for sensor and
haptic applications, whereas robust values of $\varepsilon_{33}$, and $\varepsilon_{11}$ indicate the
effectiveness as an actuator. Furthermore, notable $d_{33}$, and $d_{31}$  properties with evolution of temperature suggest that it is suitable for applications involving high temperatures,
such as sensors and haptic devices.

\subsection{Polarization}

\begin{figure}[!h]
\centering
\includegraphics[scale=0.65]{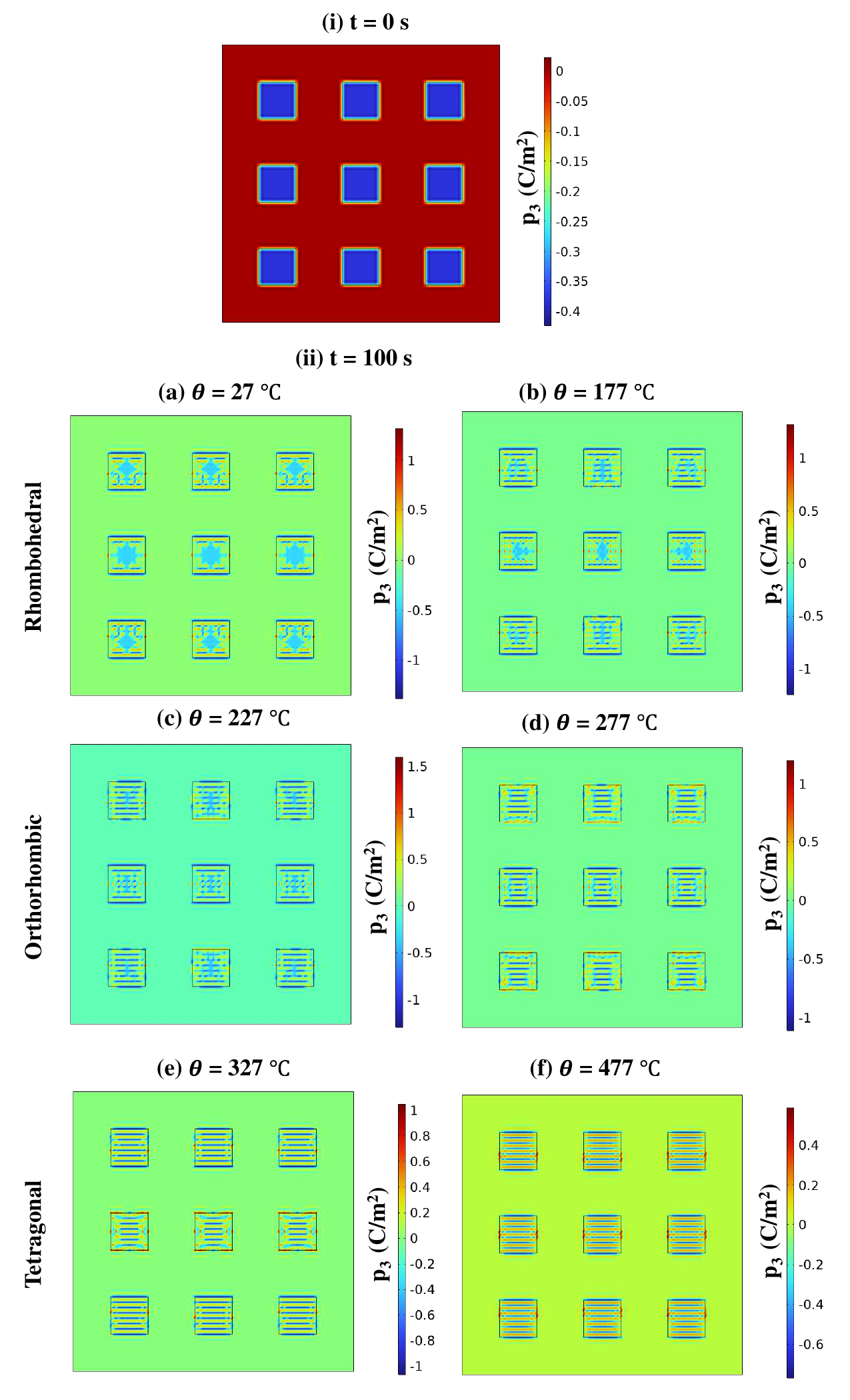}

\caption {Distribution of p$_{3}$ component of polarization.}\label{fig:P3}

\end{figure}

\begin{figure}[h!]
\centering
\includegraphics[scale=0.65]{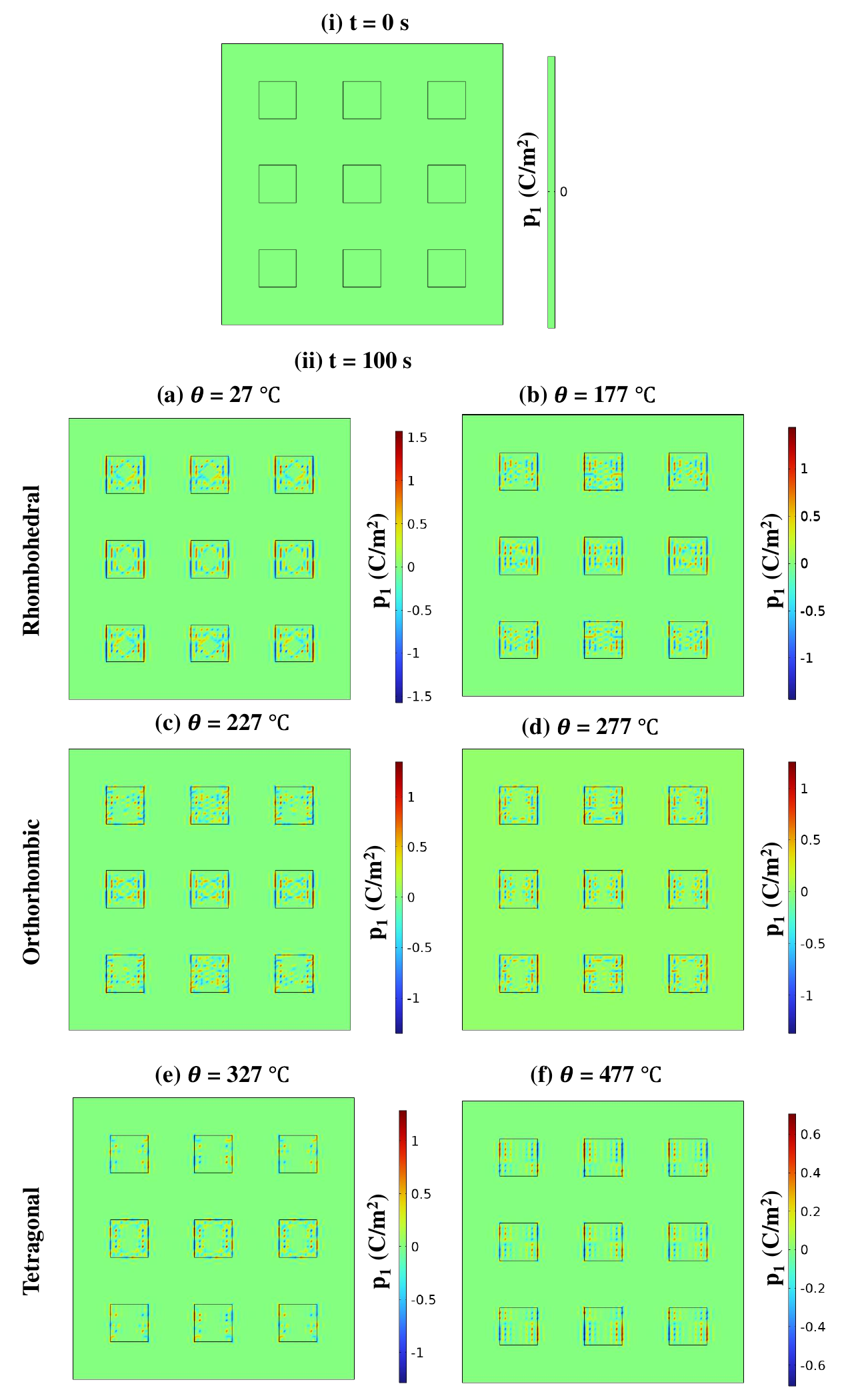}

\caption {Distribution of p$_{1}$ component of polarization.}\label{fig:P1}

\end{figure}

Figures \ref{fig:P3}, and \ref{fig:P1} exhibit the variation of different components of polarization inside the BNT-PDMS composite in $x_{3}$, and $x_{1}$ directions, respectively. The initial distribution of $p_{3}$ has negative values in the BNT inclusions because all the ferroelectric domains are lined up in the negative $x_{3}$ direction. This is because the initial electric field was in the same direction and there was no external mechanical field at the start. The PDMS matrix exhibits zero polarization because it is highly non-piezoelectric and doesn't exhibit any ferroelectricity under the studied thermal conditions. Since the initial electric field is only applied in the negative $x_{3}$ direction, there is no polarization of ferroelectric domains in the $x_{1}$ direction, and the value of the component $p_{1}$ is zero initially. With the application of an external mechanical field in the direction perpendicular to the external electric field, the distribution of both components of polarization ($p_{3}$ and $p_{1}$) changes as the applied mechanical field disturbs the initial alignment of ferroelectric domains.

The BNT inclusions exhibit twin-line patterns for $p_{3}$, in which alternate parallel lines of positive and negative values of $p_{3}$ are clearly visible. When the temperature is 27$^{\circ}C$, the BNT inclusions close to the top and bottom walls show a parallelogarm-like structure in the middle of the inclusions. They also have a higher negative value of $p_{3}$ at this location, which means that there are strongly polarized ferroelectric domains in the negative $x_{3}$ directions at this location. However, due to the close proximity of boundary walls, the parallelogram's position shifts towards the top and bottom walls, not precisely in the middle for top raw BNT inclusions and bottom raw BNT inclusions. This zone with a high negative polarization $p_{3}$ is warmer for the middle raw of inclusions and has higher negative values of $p_{3}$ for these inclusions. Because of this, these inclusions should have a higher electric potential than the top and bottom raw of BNT inclusions. Additionally, the high negative potential zone does not display a parallelogram-like structure for the middle raw inclusions, instead displaying a star-like shape. This zone exists at the exact middle position of the inclusions, which is equally affected by external conditions due to their equal distance from boundary walls. Furthermore, as we approach the middle of the inclusions, the twin-like patterns near the top and bottom walls diminish in intensity. These types of patterns are visible as 27$^{\circ}C$ is very close to room temperature, and the impact of temperature evolutions is negligible in these thermal conditions.

Furthermore, as the temperature rises to 77$^{\circ}C$, the negative polarization zone's shape and size diminish, revealing more twin-line patterns within the BNT inclusions. The negative polarization zone in the middle of BNT inclusions forms a cross-like structure, and twin lines form a recantangular pattern along the centre of this cross-shaped zone; however, the twin lines away from this zone area still alternate parallel lines, exhibiting the negative and positive values of $p_{3}$. This type of structure results from the realignment of ferroelectric domains due to temperature increases and the equal distribution of external mechanical and electric fields, with the middle raw of the BNT inclusion at an equal distance from the top and bottom boundary walls. As we move to the top and bottom raws of BNT inclusions, the left and right corner BNT inclusions exhibit diminished negative polarization zones and more twin line patterns. Due to the same magnitude and exact opposite orientation of electric potential boundary conditions at the top and bottom boundary walls, the middle inclusions of the top and bottom raws display an upward and inverted leaf-type negative zone. These inclusions are equidistance from the left and right boundary walls, where similar mechanical boundary conditions apply. This reduction in the size of the negative polarization zone is due to the weekening of intermolecular forces at high temperatures. The presence of more twin lines and a smaller negative polarization zone reduces the average value of $p_{3}$ inside the geometry. When the temperature goes up to 127$^{\circ}C$, the negative polarization's shape and size get smaller. The top and bottom raws of the BNT inclusions show a shape that looks like a middle, while the middle raw still has a thin violin-like shape. When the temperature goes up even more, to 177$^{\circ}C$, this negative polarization zone shape gets smaller, even for the left and right inclusions of the middle raw. This violin-shaped shape can only be seen at the centre inclusion of the middle raw of inclusions. The left and right inclusions have an inverted umbrella-shaped negative polarization zone because the mechanical boundary conditions at their walls are the same size but face the opposite direction. The twin-line patterns increase with an increase in temperature, which indicates further decreases in the average value of $p_{3}$. Temperature increases lead to a decrease in intermolecular forces, causing the ferroelectric domains to randomly distribute and align in a twin-line pattern under the influence of external electric and mechanical fields. The temperature of 177$^{\circ}C$ is closer to the depolarization temperature. The fact that the average value of $p_{3}$ drops near this temperature shows that ferroelectricity decreases near the depolarization temperature, making it clear that BNT becomes antiferroelectric above this temperature.

Furthermore, increasing the temperature beyond 200$^{\circ}C$ switches the phase from rhombohedral to orthorhombic via Non-ergodic relaxor (NR)/Ergodic relaxor (ER)  transition. The switching of the relaxor behavior diminishes the negative polarization zone and more twin-line patterns are observed due to high strain rate in this region. The magnitude of spontaneous polarization increases due to high strain rate, however the increase in pattern of twin lines further reduces the piezoelectric response in this phase. The sudden increase in the strain rate is due to NR/ER phase transition till 227$^{\circ}C$ and beyond that the BNT exhibits pure antiferroelectric nature, therefore the magnitude of induced spontaneous polarization decreases with increase in temperature to 277$^{\circ}C$ and beyond as shown in fig. \ref{fig:P3}. Moreover, BNT exhibits the tetragonal phase beyond 327$^{\circ}C$ and induced spontaneous polarization rises suddenly with increase in temperature and twin-line patterns intensfies even more, however, the magnitude of spontaneous polarization reduces with temperature and reduces to zero at 520$^{\circ}C$ as phase transits to cubic phase. Since the spontaneous polarization is zero in cubic phase, therefore contours have not been added to fig. \ref{fig:P3} for cubic phase. The tetragonal phase is ferroelectric in nature and therefore good values of piezoelectric coefficient is expected in this phase for BNT and BNT-based composites. The varioation of $p_{3}$ contours with temperature provide the idea that how BNT is going to behave towards thermo-electromechanical boundary conditions in different temperature ranges and it will help in deciding the approperiate usage of BNT for different type of haptic sensor and actuator problems.

Moreover, Figure \ref{fig:P1} exhibits the variation $p_{1}$ component of polarization inside the BNT-PDMS composite. The initial distribution of $p_{1}$ describes zero polarization in the $x_{1}$ direction as an electric field is applied in the $x_{3}$ direction initially. When an outside mechanical field is applied in the direction of $x_{1}$, it moves the ferroelectric domains around in all of the BNT inclusions. However, the redistribution of ferroelectric domains creates a twin-line structure in all BNT domains. The ferroelectric domains are randomly arranged in the twin lines and show a range of positive and negative values, which means that the average value of $p_{1}$ is likely to be very close to zero. The positive and negative values of $p_{1}$ along the twin lines have extreme values of more than 100$\mu C/cm^{2}$ in their respective zones, but their resultant is negligible. As the temperature rises, the intermolecular forces weaken even more, and the ferroelectric domains become less organized, even when mechanical and electric fields are applied from the outside. Based on the current contours of $p_{1}$, it is clear that it will be hard to get any electric output along $x_{1}$ for these boundary conditions. This is because the random orientation of the ferroelectric domain in this direction lessens the effect of the polarization domains. The eveolution of temperature and phases does not changes the distribution of $p_{1}$ inside the BNT inclusions significantly as shown in fig.  \ref{fig:P1} and similar type of contours are visible. Therefore, the component of polrization ($p_{1}$) in $x_{1}$ directions does not have the significant impact on the piezoelectric performance of BNT as the applied electric field is in the perpendicular diection. However, the value of $p_{1}$ varies in the similar manner as $p_{3}$ did but its amplitude is much lower compared to $p_{3}$ component of polarization. Therefore, it is suggested to have parallel direction of poling of BNT as of the electric field to have good value of piezoelectric response. The impact of both parallel and perpendicular poling on various parameters such as mechanical strain and electric potential is further discussed in the later section of the paper.
\subsection{Electric potential distribution}
\begin{figure}[h!]
\centering
\includegraphics[scale=0.65]{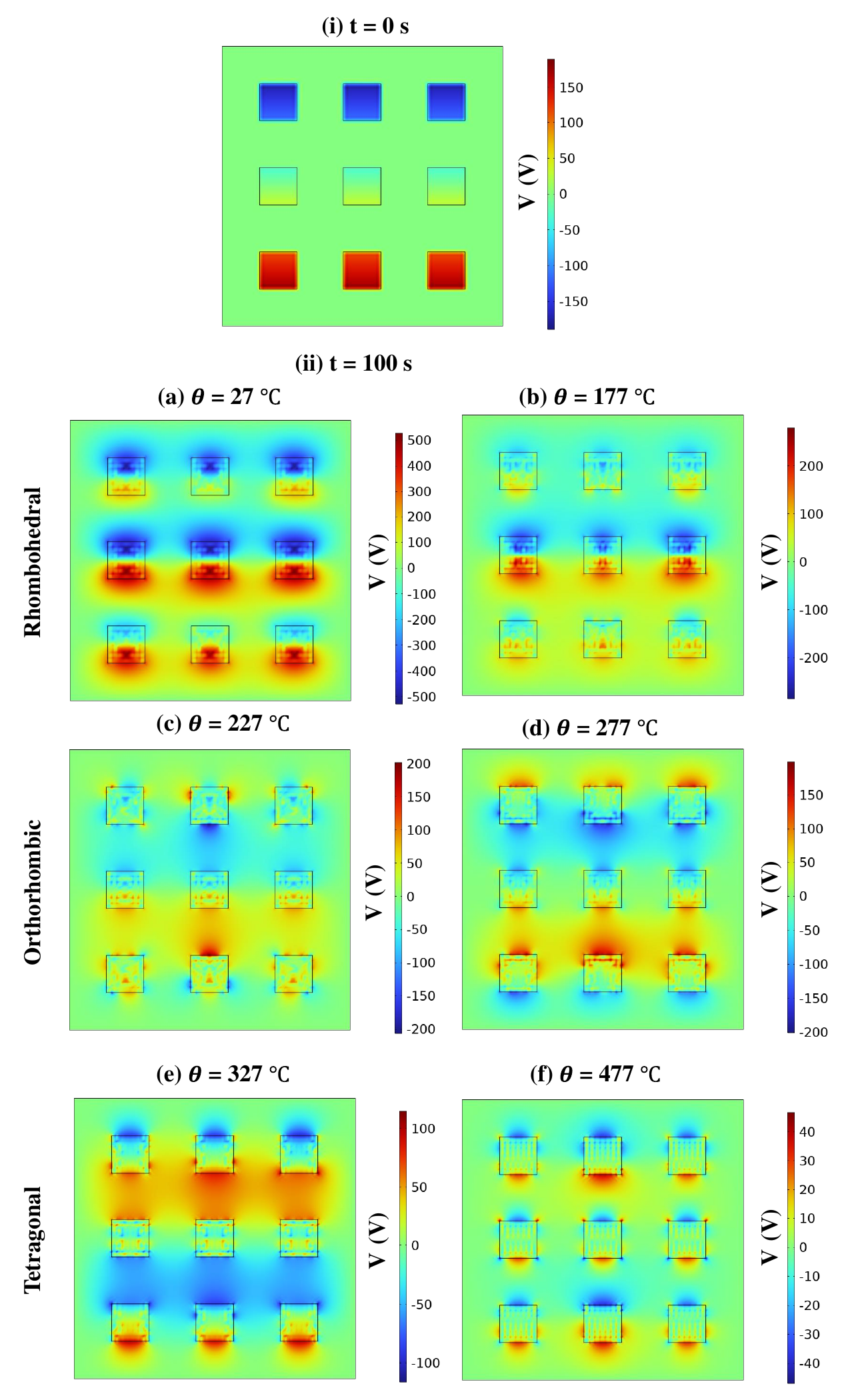}

\caption {Electric potential distribution (V)}\label{fig:V}

\end{figure}

Figure \ref{fig:V} displays the spatial arrangement of electric potential within the BNT-PDMS composite. The initially supplied electric field is antiparallel to the spontaneous polarization, as seen in Fig. \ref{fig:V}. Nevertheless, the electric potential's magnitude is rather weak due to the fact that the ferroelectric domains are only aligned in the negative $x_{3}$ direction. Applying a mechanical field disrupts the distribution of ferroelectric domains and causes twin-line patterns with a diamond-shaped negative polarization zone to form at the centre of BNT inclusions. This leads to an increase in the avalanche spontaneous polarization ($p_{3}$) of the inclusions, which in turn increases the value of the avalanche electric potential and its distribution along the BNT inclusions. The electric potential generated along the middle row of BNT inclusions is greater than that of the top and bottom rows due to the higher negative zone of spontaneous polarization at the middle row of BNT inclusions at a temperature of 27 $^{\circ}C$. As the temperature climbs to 77$^{\circ}C$, the negative spontaneous polarization zone shrinks while the twin-line zones expand. Consequently, the average values of electric potential fall. Nevertheless, the electric potential distribution surrounding each BNT inclusion adheres to the outlines of spontaneous polarization ($p_{3}$) inside the BNT-PDMS composite. As previously mentioned, the influence of spontaneous polarization ($p_{1}$) on the characterization of ferroelectric domains is insignificant, and hence it also has a minor effect on the distribution of electric potential. The insertion of BNT inclusions between the top and bottom layers has resulted in a wider range of negative spontaneous polarization zone, leading to a higher distribution of electric potential on the subsequent BNT inclusions. The size of the left and right side inclusions of the top and bottom raw inclusions is reduced because of the diverse range of twin-line patterns in the spontaneous polarization contours. The electric potential gradient is highest at the centre BNT inclusion of the BNT-PDMS composite because it has the largest negative spontaneous polarization zone. This polarization zone generates the most transformational strain because the ferroelectric domains align in the negative $x_{3}$ direction. However, the distribution of elastic strain is uniform across all the inclusions.

Moreover, when the temperature increases further, the electric potential gradient decreases even further. This occurs due to the contraction of the negative spontaneous polarization region, resulting in the misalignment of ferroelectric domains. As a result, the material undergoes a shift from being ferroelectric to being antiferroelectric.The electric potential contours at 127 $^{\circ}C$ show a decrease in magnitude compared to those at 77 $^{\circ}C$. However, the difference is not as significant as it was at 177 $^{\circ}C$, as this temperature is closer to the depolarization temperature and the NR/ER boundary. This phenomenon refers to the decrease in intermolecular interactions as temperature increases, resulting in the alignment of ferroelectric domains in a random manner. As temperature continues to rise, the degree of antiferroelectricity increases. Furthermore, the electric potential distribution along each BNT inclusion exhibits a similar pattern to that of the spontaneous polarization ($p_{3}$). This pattern consists of a maximum gradient at the middle inclusions of each row of BNT inclusions, and a minimum gradient of electric potential at the left and right inclusions of the top and bottom rows of BNT inclusions. The central inclusion of the BNT-PDMS composite displays the highest electric potential gradient.

The electric potential contours shown in Fig. \ref{fig:V} correspond to a temperature range of 27 $^{\circ}C$-177$^{\circ}C$, specifically inside the rhombohedral phase. However, when the temperature increases, the shift in the patterns of spontaneous polarization and electric potential distribution indicates that the lattice structure becomes disoriented from its strong rhombohedral structure. However, it is expected to undergo a full transformation into an orthorhombic structure at a temperature of 200$^{\circ}C$, as a result of a phase shift and the dynamics of micro and nano domains. This transformation will have a significant effect on the material's piezoelectric response. The investigation of the electromechanical behaviour of BNT at temperatures over 200 $^{\circ}C$ is intended and expected to be carried out as a future expansion of the ongoing work. The electric potential contours within the temperature range studied (27 $^{\circ}C$-177$^{\circ}C$) indicate that BNT demonstrates favourable electric potential values as an output, even when subjected to lower external mechanical strains. This makes it a viable choice for use as a sensor in haptic and other applications. Nevertheless, the performance of BNT declines as the temperature rises beyond 200$^{\circ}C$ due to NR/ER transition and complete reversibilty of polarity of electric potential is observed in this transition and orthorhombic phase and dinished magnitude of electric potential are observed at 227$^{\circ}C$ and which reduces further at 277$^{\circ}C$ due to purely antiferroelectric nature of BNT in orthorhombic phase. Moreover, increasing the temperature beyond 320$^{\circ}C$, changes the phase to tetragonal phase and increases the electric potential gradient across the BNT inclusions due to ferroelectric nature. The magnitude of electric potential gradient is going to decrease with increase in temperature due to decreases in the spontaneous polarization with temperature in this phase and lower value of electric potential is observed at 477$^{\circ}C$. However, BNT still demonstrates high electric potential values within this temperature range. Additionally, it is important to note that BNT will not exhbit any value of electric potential generation beyond 520$^{\circ}C$ due to zero polarization in cubic phase, therefore contours beyond 520$^{\circ}C$ is not included in Fig. \ref {fig:V}. Moreover, BNT exhibits lower values of electric potential generation at temperature of 477$^{\circ}C$ but it is still quite significant and even more significant in all early temperatures of tetragonal phase which makes it a suitable choice of high temperature haptic sensors and actuator applications. Furthermore, some modifications regarding composition and designs of BNT-based piezoelectric is reguired to enhance the electric potential output at high temperatures . This is a promising area of future research based on the current work. 
\subsection{Strain distribution}
\begin{figure}[h!]
\centering
\includegraphics[scale=0.65]{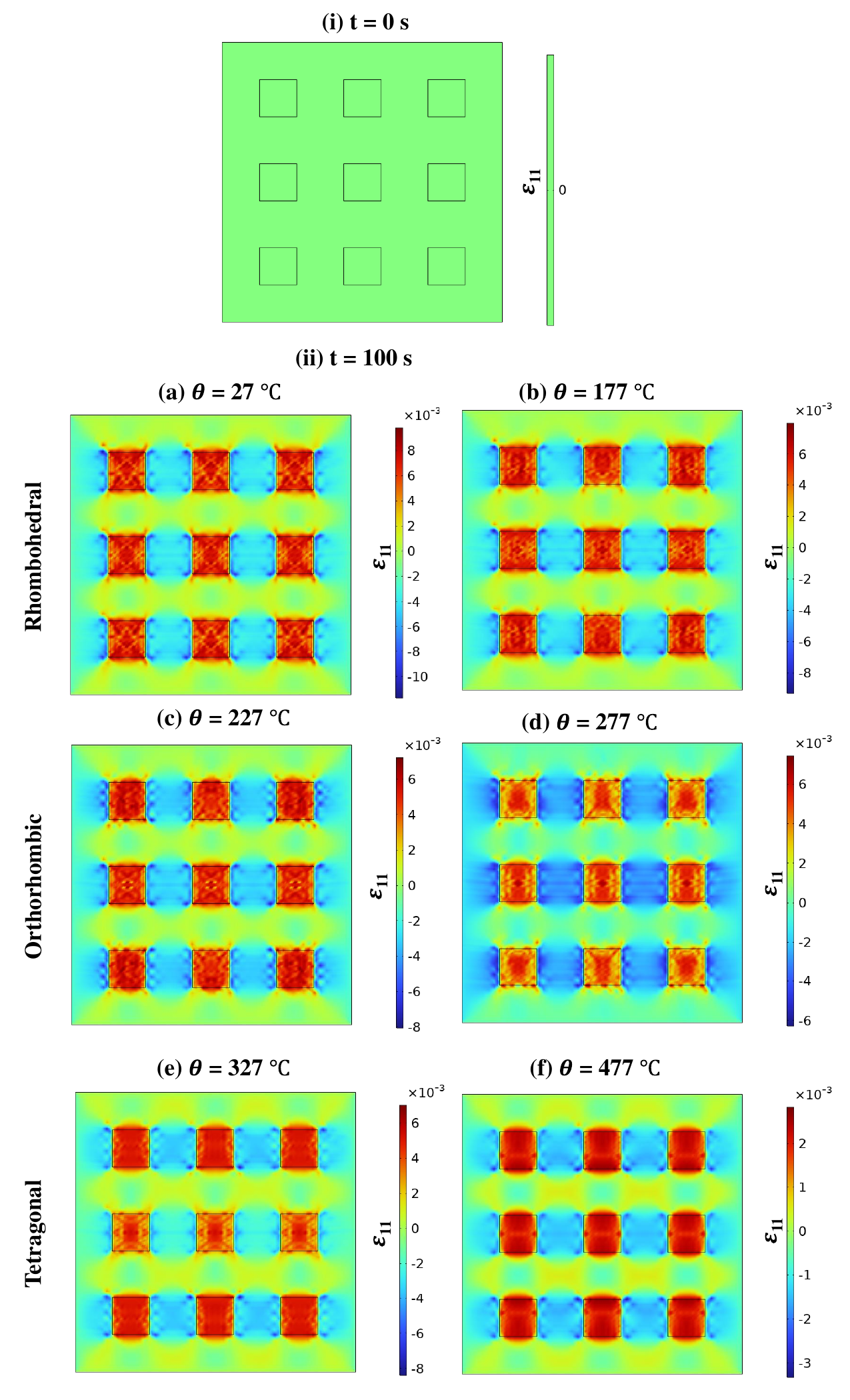}

\caption {Principal strain distribution ($\varepsilon_{11}$).}\label{fig:u1x}

\end{figure}

\begin{figure}[h!]
\centering
\includegraphics[scale=0.65]{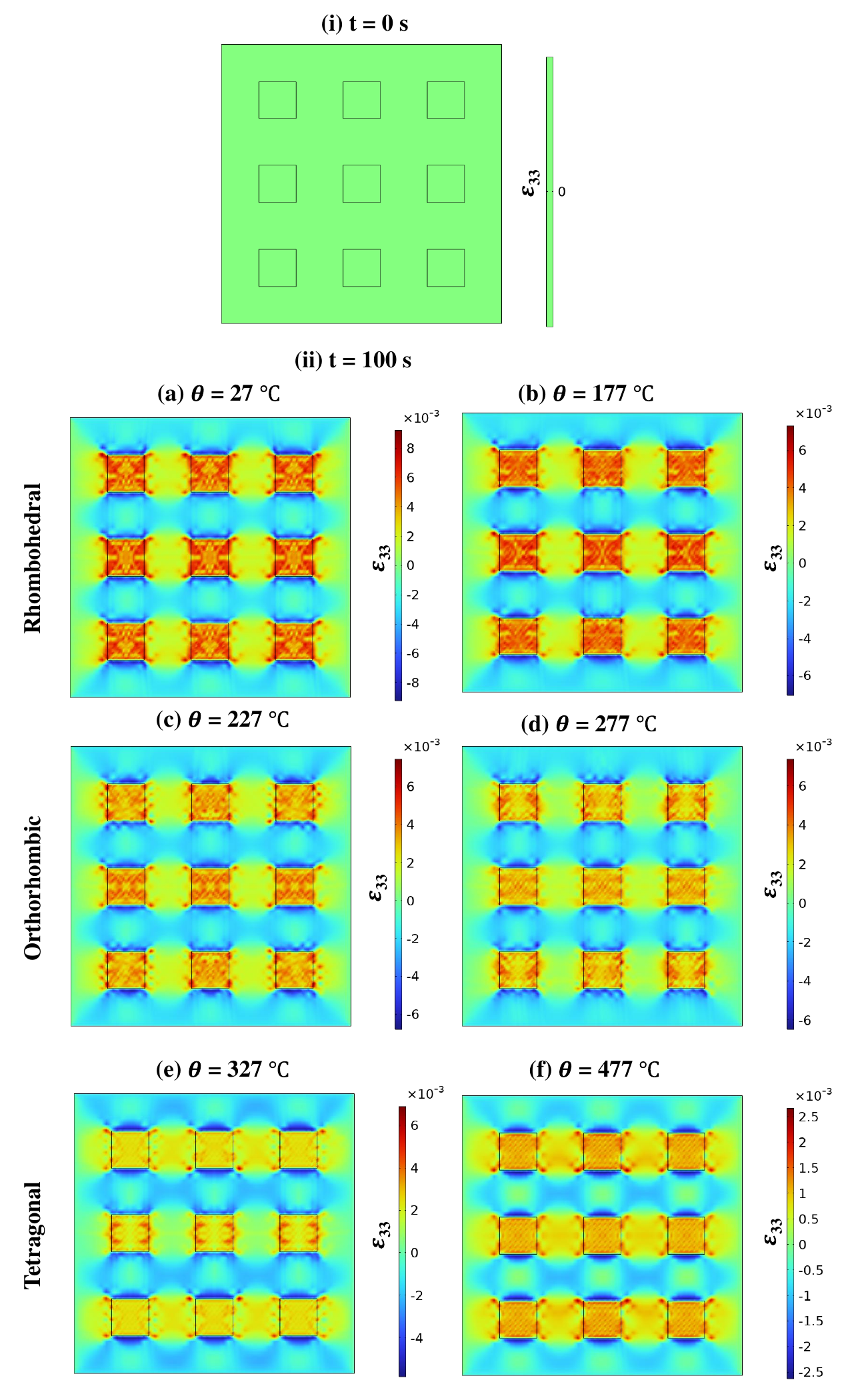}

\caption {Principal strain distribution ($\varepsilon_{33}$).}\label{fig:u3y}

\end{figure}

Figures \ref{fig:u1x} and \ref{fig:u3y} display the spatial arrangement of the primary strains $\varepsilon_{11}$ and $\varepsilon_{33}$ along the $x_{1}$ and $x_{3}$ axes, respectively, inside the BNT-PDMS composite. The total mechanical strain comprises both the elastic strain and transformation strain. The elastic strain arises from the development of elastic stress, while the transformational strain is a result of the spontaneous polarization of ferroelectric domains. Although the contribution of transformation strain is not as significant as that of elastic strain, it has a notable impact on the dielectric response of piezoelectric materials. This effect is clearly observed in the enhanced generation of electric potential.

At the beginning, no external mechanical strain field is given to the BNT-PDMS composite, resulting in the absence of any primary strains ($\varepsilon_{11}$ and $\varepsilon_{33}$) as seen in the figures. The figures \ref{fig:u1x} (i) and \ref{fig:u3y} (i) correspondingly. The elastic strain in the entire BNT-PDMS composite is uniform when a uniform mechanical strain is applied in the $x_{1}$ direction. However, the BNT inclusions develop a positive transformation strain due to spontaneous polarization, which induces a precisely negative strain in the PDMS composite in the $x_{1}$ direction for $\varepsilon_{11}$ and in the $x_{3}$ direction for $\varepsilon_{33}$. The amplitude of transformation strain in BNT inclusions follows a similar pattern to the variation in spontaneous polarization for the given distribution of BNT inclusions. The primary cause of the transformation strain is the spontaneous polarization $p_{3}$ in $\varepsilon_{33}$, while the impact of $p_{1}$ in $\varepsilon_{11}$ is insignificant. Consequently, the transformation strain in the $x_{1}$ direction is also influenced by $\varepsilon_{33}$ as the lateral strain and longitudinal strain are interconnected through the Poisson's ratio.

The $\varepsilon_{33}$ within BNT inclusions display a star-like structure, as depicted in Fig. \ref{fig:u3y} (ii). Conversely, a proportional negative strain is induced in the $x_{3}$ direction within the PDMS matrix due to transformational strain. The mechanical strain in the $x_{1}$ direction is primarily a result of an externally applied mechanical field in the same direction. Consequently, positive values of $\varepsilon_{33}$ are observed alongside the inclusions in the $x_{1}$ direction. The magnitude of $\varepsilon_{33}$ in BNT inclusions varies within the inclusion and resembles a cavity with a negative polarization zone at 27 $^{\circ}C$. This cavity has a lower strain magnitude compared to the wings of the star, resulting in a strain gradient within the inclusion. Consequently, this strain gradient generates a high electric potential. The morphology of this cavity changes as one moves towards the upper and lower rows of BNT inclusions, in response to variations in the direction and amplitude of the applied mechanical and electric fields. The patterns on the top row of BNT inclusions are a precise mirror copy of the patterns on the bottom row of BNT inclusions. Additionally, the geometry of these cavities is precisely comparable but inverted. In addition, when the temperature rises, the value of $\varepsilon_{33}$ lowers, and the size of these cavities also reduces due to the emergence of twin-line patterns in the ferroelectric domains at higher temperatures in rhombohedral phase. Consequently, a drop in the piezoelectric response is observed at higher temperatures due to a reduction in both the amplitude and distribution of strain gradient as the temperature increases.  Moreover, the strain gradient inside the BNT inclusion increases suddenly due to NR/ER tarnsition beyond 200$^{\circ}C$, which is clearly visible at 227$^{\circ}C$ but the $\varepsilon_{33}$ decreases with temperature and shows lower values at higher temperature such as 277$^{\circ}C$ in orthorhombic phase. However, this strain $\varepsilon_{33}$ is higher than the $\varepsilon_{33}$ in tetragonal phase as BNT exhibits antiferroelectric nature in orthorhombic phase while ferroelectric nature in tetragonal phase. With the evolution of temperature, the $\varepsilon_{33}$ again decreases as intermolecular forces decreases with rise in the temperature which will further decreases the piezoelectric performance as BNT approaches the Curie temperature as ploarization diminishes after this point and only elastic strain will be there and transformation strain will be zero. However, the performance of BNT is better in tetragonal phase which suggests its usage in high temperature haptic sensor and actuator applications.

Furthermore, Fig. \ref{fig:u1x} (ii) demonstrates the fluctuation of $\varepsilon_{11}$ in response to variations in temperature and the application of external mechanical strain. The distribution of transformational strain for $\varepsilon_{11}$ within BNT inclusions is uniform because of the negligible average spontaneous polarization $p_{1}$ and the existence of alternating vertical twin-line patterns of $p_{1}$. The strain gradient is characterized by a small magnitude, and the transformational strain generated by $\varepsilon_{33}$ remains relatively constant in the inclusions and corners of BNT inclusions. By applying an external mechanical strain field in the $x_1$ direction, it is possible to see negative values of $\varepsilon_{11}$ near inclusions in the same direction. Furthermore, there are comparable positive values in the $x_3$ direction within the PDMS matrix. The negative correlation between lateral strain and longitudinal strain is attributed to the Poisson ratio. The impact of $\varepsilon_{11}$ is low since the average $p_1$ is small and an external electric field is applied in the $x_3$ direction (BNT is polarized in the $x_3$ direction). Hence, the electric potential needed is in the $x_3$ direction, and the current distribution of $\varepsilon_{11}$ sustains the electric potential in the $x_3$ direction. Moreover, the magnitude of the transformative aspect of $\varepsilon_{11}$ decreases with increasing temperature as a result of a decrease in intermolecular interactions, leading to a reduced ability to resist deformation. Moreover, $\varepsilon_{11}$ exhibits the similar performance as of $\varepsilon_{33}$ in orthorhombic and tetragonal phases with the evolution of temperature. However, the impact of transformation strain is not that significant but still elastic strain also decreases with increase in temperature due to weakening of intermolecular forces. The strains $\varepsilon_{11}$ and $\varepsilon_{33}$ exhibits good values in both the directions $x_{1}$ and $x_{3}$ directions respectively, however the electric ouput exhibits lower values in $x_{1}$ direction than $x_{3}$ direction. Therefore, further research is required to modify the composition and design of BNT-based piezoelectric composites for better piezoelectric performance.
The strain distribution contours depicted in Figs. \ref{fig:u1x} and \ref{fig:u3y} pertain to a temperature range of 27 $^{\circ}C$-177$^{\circ}C$, specifically inside the rhombohedral phase. However, when the temperature rises, the change in the patterns of spontaneous polarization and mechanical strain distribution suggests that the lattice structure loses its alignment with the strong rhombohedral structure. Nevertheless, it is anticipated that the material would completely change into an orthorhombic structure when exposed to a temperature of 200$^{\circ}C$, due to a phase transition and the movement of micro and nano domains. This change will have a substantial impact on the material's transformational strain and, consequently, its piezoelectric response. There are plans to conduct an inquiry into the electromechanical properties of BNT at temperatures over 200 $^{\circ}C$ as part of future research.
\subsection{Effective piezoelectric coefficients}
\begin{figure}[h!]
\centering
\includegraphics[scale=0.55]{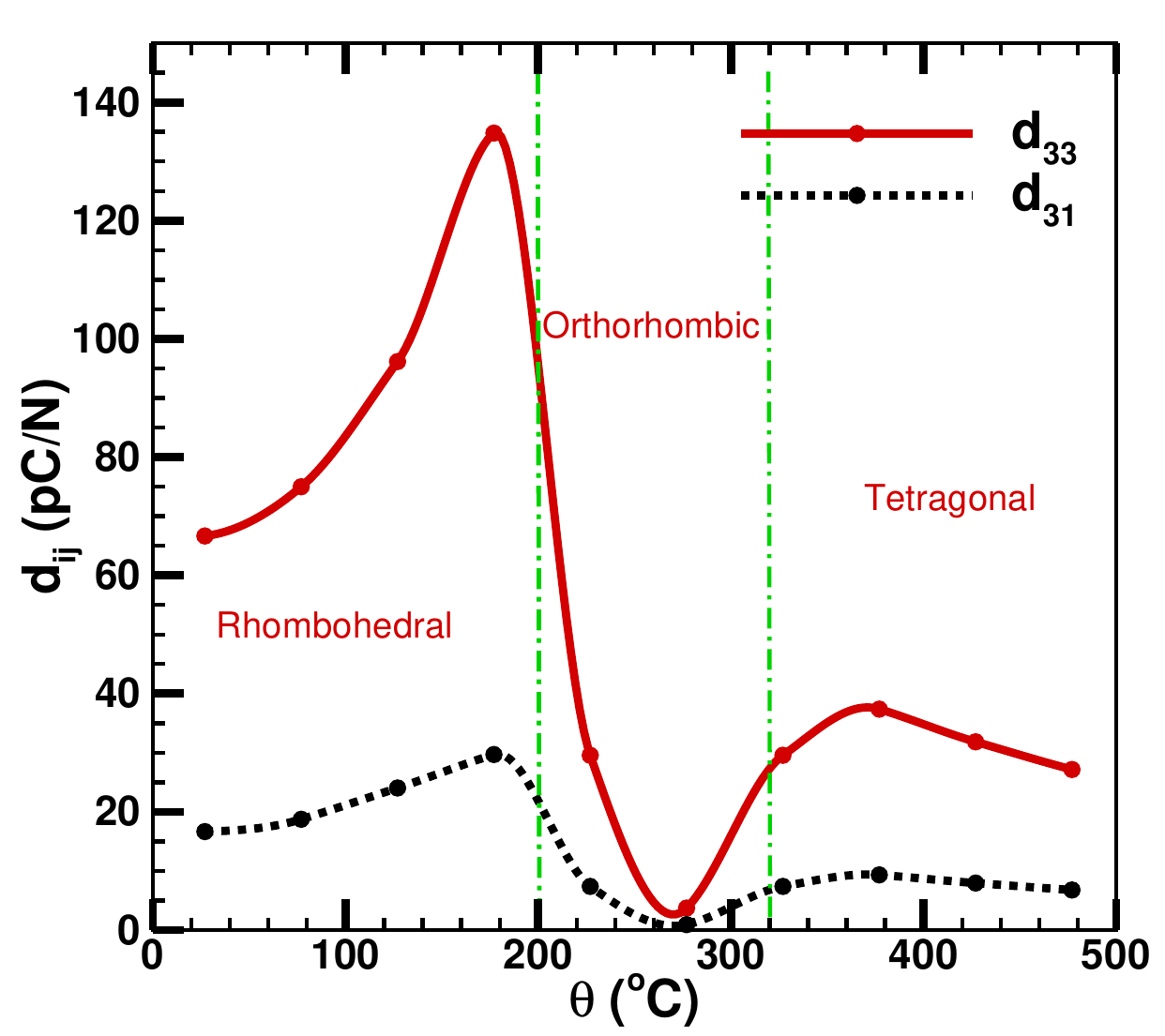}

\caption {Effective piezoelectric coefficients (d$_{33}$, and d$_{31}$) induced by spontaneous polarization.}\label{fig:d}

\end{figure}

Figure \ref{fig:d} illustrates the relationship between the change in effective piezoelectric coefficients with the temperature's progression, caused by spontaneous polarization. The piezoelectric coefficients $d_{33}$ and $d_{31}$ represent the rates of change of strain, $\varepsilon_{33}$ and $\varepsilon_{11}$, respectively, when an electric field is absent and the strain is applied parallel or perpendicular to the switching directions.  The piezoelectric coefficients ($d_{33}$ and $d_{31}$) establish the connection between the strain produced and the electric field applied. They may be seen as a measure of the sensitivity to force, specifically the amount of charge released per Newton of force. However, there are external null stresses but external applied strain and electric field induces the resistive strain and stress inside the composite which in turn induces electric potential due to electromechanical coupling. Additionally, there will be transformational strain due to ferroelectric polarization which will also induces the non-linear effects in strain and electric potential generation.
The values of $d_{31}$ and $d_{33}$ are supposed to increase with the progression of temperature irrespective of their directions. However, the overall value of $d_{31}$ is smaller compared to $d_{33}$ as both electric and mechanical field are perpendicular to each other and effect of applied mecahnical strain in $x_{1}$ direction which is the induced transverse strain in $x_{3}$ direction is smaller for same applied field, however; for the case of $d_{33}$ both applied mechanical strain and electric field are in same direction so giving a little bitt higher values. The piezoelectric coefficients exhibit lower values at room temperature which increases linearly with increase in temperature, which is due to increase in strain responsiveness with temperature as intermolecular forces decreases and produce increased mechanical strain. We also can elobrate that lesser amount of mechanical force is required to produce same mechanical strain and to release same amount of charge. The piezoelectric coefficients increases being the ratio of charge produced per newton of force. The variation of piezoelctric coefficients is linear in rhombohedral phase i.e till depolarization temperature but the curve changes its shape and slope suddenly after depolarization temperature and starts decreasing with temperature due to NR/ER transition as ferroelectric domain transists to antiferroelectric nature and completly transforms to antiferroelectric nature at 227 $^{\circ}C$ and further keep on reducing in orthorhombic phase and gets a minimum value at 277 $^{\circ}C$ due to fully antiferroelectric domains in orthorhombic phase. The relaxor behaviour of BNT is charcterized by change of slope and curvature of $d_{ij}-\theta$ curve due to changes of phase boundary from NR which is ferroelectric to ER which is antiferroelectric on nature. The slope of $d_{ij}-\theta$ curve becomes negative and curve again become linear as relaxor behavior changes completely to ER and no or very few domain switching is observed in this phase. After 277 $^{\circ}C$, BNT starts to transit from orthorhombic to tetragonal phase and increases with increase in temperature in tetragonal phase till 377 $^{\circ}C$ exhibiting a local maxima and start reducing after that as spontaneous polarization decreases with increase in temperature. The $d_{ij}-\theta$ curve beyond 477 $^{\circ}C$ has not been shown in  Fig. \ref{fig:d} as in the cubic phase spontaneous polarization becomes zero and BNT exhibits the dielectric nature of domains in the cubic phase.  Moreover, the overall piezoelectric coefficient for BNT seems to exhibit the same variation with temperature irrespective of switching directions, however; the magnitude of $d_{31}$ is much lower compared to $d_{33}$ due to parallel poling directions of electric and mechanical fields in the later case. Furhthermore, the enhancement of performance of BNT-based piezoelectric compoistes are also required by modifying the compositions and designs of BNT inclusions inside the composites. 

\subsection{Effective elastic coefficients}

\begin{figure}[!h]
\centering
\includegraphics[scale=0.55]{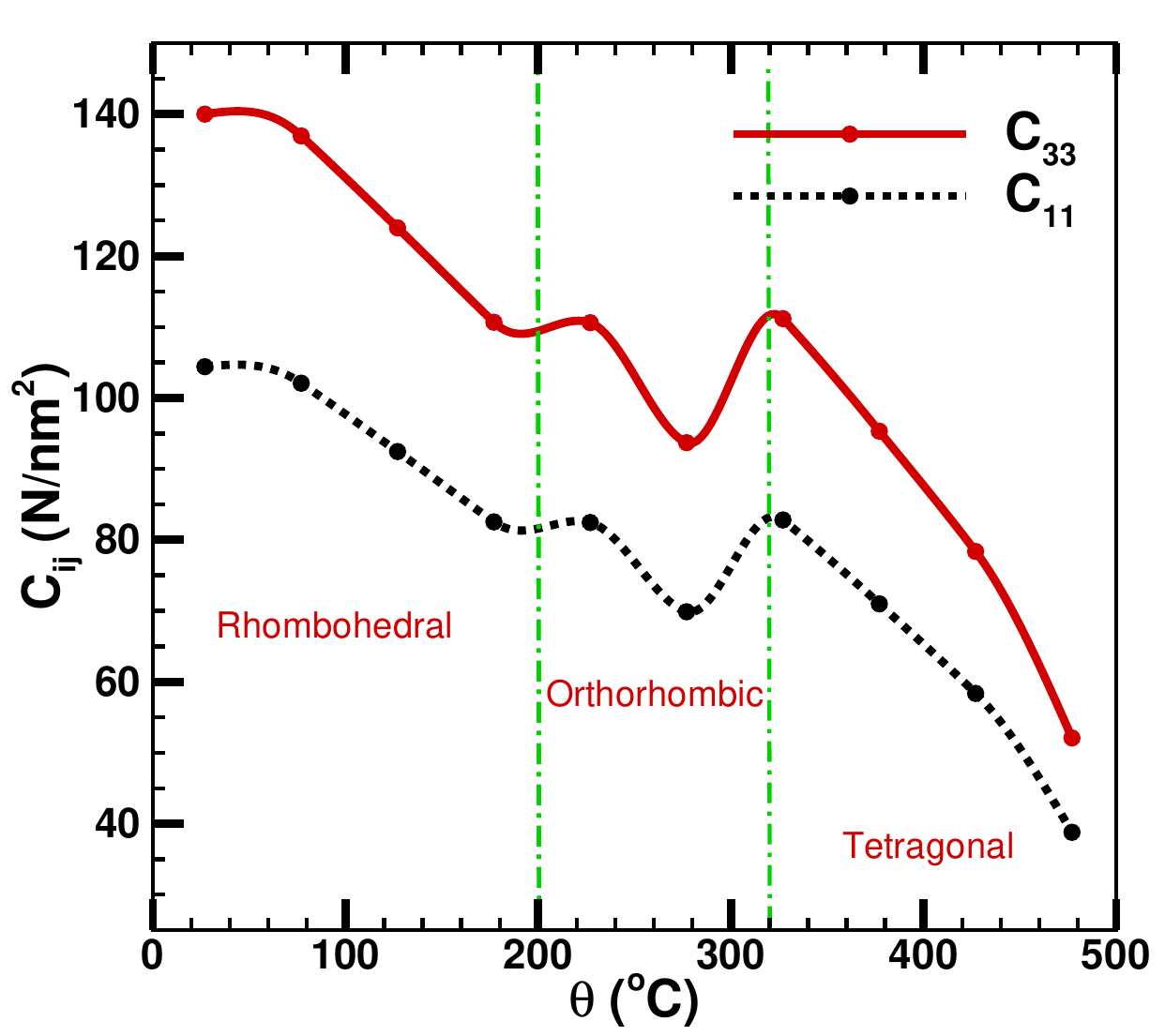}

\caption {Effective elastic coefficients (C$_{33}$, and C$_{11}$) induced by spontaneous polarization.}\label{fig:C}

\end{figure}

Figure \ref{fig:C} illustrates the relationship between the change in effective elastic coefficients with the temperature's progression, caused by spontaneous polarization. The elastic coefficients $C_{33}$ and $C_{11}$ represent the rates of change of normal stresses, $\sigma_{33}$ and $\sigma_{11}$, with respect to total mechanical strains $\varepsilon_{33}$ and $\varepsilon_{11}$ respectively, when an electric field is absent and the strain is applied parallel or perpendicular to the switching directions.  The elastic coefficients ($C_{33}$ and $C_{11}$) exhibits the resilience shown by BNT-PDMS composite towards the externally applied mechnical field. They may be seen as a measure of the ability to resist theexternal applied mechanical force. 
The values of $C_{33}$ and $C_{11}$ decreases with temperature in the rhombohedral phase as the intermolecular forces decreases with increase in temperature and lesser mechanical force is required to get the same deformation at higher temperature till depolarization temperature. After depolarization temperature, the relaxor behaviour switches from NR to ER relaxor behaviour which transists the parallel ferroelectric domains to antiferroelectric domains and transformational strain increases with increase in temperature, the $C_{ij}-\theta$ curve becomes stangant at 177 $^{\circ}C$ and remains stangant till 227$^{\circ}C$. BNT completely transists to antiferroelectric orthorhombic phase after 227 $^{\circ}C$ and $C_{ij}-\theta$ curve starts to decrease and reaches a local minima at 277 $^{\circ}C$. Thereafter, the transition from orthorhombic phase to tetragonal phase begins and the values of both $C_{11}$ and $C_{33}$ increases suddenly and reaches a maximum value at 327$^{\circ}C$ and then starts to decrease with temperature as spontaneous polarization decreases with temperature. The curvature of $C_{33}$ and $C_{11}$ are very similar, however, the magnitude of $C_{11}$ is lower than $C_{33}$ as BNT exhibits higher elastic properties in $x_{3}$ direction than the $x_{1}$ direction. The higher value of $C_{33}$ is an indiaction of higher mechanical responsiveness in $x_{3}$ direction. Therefore, it is suggested to apply external mechanical field in $x_{3}$ direction which means parallel application of electrical and mechanical field or poling in $x_{3}$ direction for better piezoelectric response and strain responsiveness. Moreover, further modifications in compositions and design of BNT-based composites to enhance the piezoelctric performance.

\section{Conclusions}

A two-dimensional computational model has been developed to analyze the influence of phase-field thermoelectromechanical performance on the lead-free BNT-PDMS composite under various temperature haptic stress conditions. A thermoelectromechanical model based on phase-field analysis investigates the phase transition and micro-domain dynamics of BNT-based piezoelectric materials and their effects on piezoelectric performance and strain responsiveness. Due to its intricate phase and domain structures, which are fully dependent on the material's temperature and become even more complex at higher temperatures, the thermal stability of such material must be assessed. Under different electric and mechanical boundary conditions, the effective elastic and piezoelectric characteristics, as well as the mechanical and electric field parameters, of the two-dimensional framework have been examined. From the results of the analyse, several deductions may be made.
\begin{itemize}
\item The distribution of spontaneous polarization $p_{3}$ inside the BNT inclusions mimics twin-line patterns with an enlarged negative polarization zone in the center of BNT inclusion in rhombohedral phase, which diminishes with increase of temperature and completely vanishes in orthorhombic and tetragonal phases. This indicates that BNT must have high piezoelectric performance in rhombohedral phase instead of any other phase.
\item The reversal of polarity of electric potential distribution in orthorhombic phase confirms the antiferroelectric nature of BNT in this phase. The electric potential generation is highest in rhombohedral phase which decreases in orthorhombic phase. The electric potential gradient again increases in tetragonal phase which further decreases with evolution of temperature as spontaneous polarization decreases with evolution of temperature. BNT exhibits the dielectric nature in cubic phase as spontaneous polarization vanishes to zero at Curie temperature.
\item BNT exhibits highest strain gradients during NR/ER transition due to reversal of ferroelectric domains to antiferroelectric in this temperature range, however the mechanical strain is higher in orthorhombic phase due to ferroelectric domains. The high strain gradients in this phase exhibits that BNT has great strain responsiveness in this phase and very much suitable for actuator applications. The overall mechanical strain decreases with increase of temperature in both rhombohedral and tetragonal phases.
\item Effective piezoelectric coefficients increase with increase in temperature in rhombohedral phase, which is obvious with increase in temperature the ferroelectric domains allow more electric output to generate even with small application of mechanical field. However, after depolarization temperature, ferroelectric domains starts switching to antiferroelectric which suddenly decreases the effective piezoelectric coefficient due to NR/ER transition which further decreases in orthorhombic phase. The effective piezolectric coefficient again increases as BNT transists to tetragonal phase but with evolution of temperature, it decreases as spontaneous polarization decreases with temperature.
\item Effective elastic coefficients also exhibit decrease in magnitude with evoultion of temperature in both rhombohedral and tetragonal phases as intermolecular forces decreases and all domains are parallely alligned in ferroelectric domains. However, BNT exhibits a constant stiffness during NR/ER transition due to resistance creasted by ferroelectric to antiferroelectric switching, thereafter the values again decreases with temperature as BNT turns fully antiferroelectric in orthorhombic phase. The values again increase due to orthorhombic to tetragonal ohase transition and then decreases with increase in temperature in tetragonal phase. The resilience in NR/ER transition phase exhibits its strain responsiveness in that phase and make it very suitable for the application where higher mechanical output is desired.
\item The higher values of $d_{33}$ and $C_{33}$ compared to the values of $d_{31}$ and $C_{11}$ exhibits that BNT exhibits better performance when externally applied mechanical and electric field are applied in $x_{3}$ direction which means parallel or antiparallel poling compared to perpendicular poling of BNT-PDMS composite.
\end{itemize}

It is observed that the current phase-field thermoelectromechanical model is able to mimic the phase change and micro-dynamics behaviour of BNT-based composite. The impact of temperature and phase change on the performance of BNT-PDMS composite is significant and can not be ignored. Further, the findings of this paper forms the basis of further performance enhancement of BNT-based piezoelectric by design and composition modification of the composites and piezolelctric. The PDMS matrix is soft and non-piezoelctric in nature, therefore addition of carbon nano-tubes (CNT) can clearly enhance the mechanical and electrical durability of BNT-PDMS composite, which can also effect micro-domain dynamics of BNT. Therefore it can be a future aspect of the current study. Moreover, the shape and distribution of BNT inclusions can also affect the microdynamics of domains and thus the piezoelctric performance, hence; it could also be a potential basis of future study in this field.

\section*{Acknowledgements}

The authors are grateful to the NSERC and the CRC Program (Canada) for their support. This publication is part of the R$^{+}$D$^{+}$i project, PID2022-137903OB-I00, funded by MI-CIU/AEI/10.13039 /501100011033/ and by ERDF/ EU. This research was enabled in part by support provided by SHARCNET (www.sharcnet.ca) and Digital Research Alliance of Canada (www.alliancecan.ca).

\appendix
\section{Minimization strategies for different phases}\label{A}
\subsection{3D case}
In order to minimize $W^{3D}(\theta,\pmb{\overrightarrow{p}})$, we need to minimize the Helmholtz free energy function with respect to $\pmb{\overrightarrow{p}}$ and we will get:
\begin{equation}
\phi_{\pmb{\overrightarrow{p}}} = W^{3D}_{\pmb{\overrightarrow{p}}}(\theta,\pmb{\overrightarrow{p}})-\pmb{\overrightarrow{E}} = 0. \label{A1}
\end{equation}
Now all three components of electric field can be written as:
\begin{equation}
E_{1} = 2\alpha_{1}\frac{\theta_{c}-\theta}{\theta_{c}}p_{1}+4\alpha_{11}p_{1}^{3}+2\alpha_{12}\frac{\theta_{c}-\theta}{\theta_{c}}p_{1}(p_{2}^{2}+p_{3}^{2})+6\alpha_{111}p_{1}^{5}, \label{A2}
\end{equation}
\begin{equation}
E_{2} = 2\alpha_{1}\frac{\theta_{c}-\theta}{\theta_{c}}p_{2}+4\alpha_{11}p_{2}^{2}+2\alpha_{12}\frac{\theta_{c}-\theta}{\theta_{c}}p_{2}(p_{2}^{2}+p_{1}^{2})+6\alpha_{111}p_{2}^{5}, \label{A3}
\end{equation}
\begin{equation}
E_{3} = 2\alpha_{1}\frac{\theta_{c}-\theta}{\theta_{c}}p_{3}+4\alpha_{11}p_{3}^{2}+2\alpha_{12}\frac{\theta_{c}-\theta}{\theta_{c}}p_{3}(p_{1}^{2}+p_{2}^{2})+6\alpha_{111}p_{3}^{5}. \label{A4}
\end{equation}
In absence of electric field, the solution of the Eqs. (\ref{A2}), (\ref{A3}), and (\ref{A4}) can be written as:
\begin{equation}
p_{1} =0, \alpha_{1}\frac{\theta_{c}-\theta}{\theta_{c}}+2\alpha_{11}p_{1}^{2}+\alpha_{12}\frac{\theta_{c}-\theta}{\theta_{c}}(p_{2}^{2}+p_{3}^{2})+3\alpha_{111}p_{1}^{4} = 0, \label{A5}
\end{equation}
\begin{equation}
p_{2} =0, \alpha_{1}\frac{\theta_{c}-\theta}{\theta_{c}}+2\alpha_{11}p_{2}^{2}+\alpha_{12}\frac{\theta_{c}-\theta}{\theta_{c}}(p_{3}^{2}+p_{1}^{2})+3\alpha_{111}p_{2}^{4} = 0, \label{A6}
\end{equation}
\begin{equation}
p_{3} =0, \alpha_{1}\frac{\theta_{c}-\theta}{\theta_{c}}+2\alpha_{11}p_{3}^{2}+\alpha_{12}\frac{\theta_{c}-\theta}{\theta_{c}}(p_{1}^{2}+p_{2}^{2})+3\alpha_{111}p_{3}^{4} = 0. \label{A7}
\end{equation}
To characterize the first-order phase transition, we use the assumption that $\alpha_{111}$ and $\alpha_{12}$ are both positive, $\alpha_{11}$ is negative, and $\alpha_{1}\frac{\theta_{c}-\theta}{\theta_{c}}$ varies with temperature, changing sign when the transition temperature is reached. When the value of $\alpha_{1}\frac{\theta_{c}-\theta}{\theta_{c}}$ is negative, the minimum of the free energy will result in polarizations of a limited magnitude. In Eqs.  (\ref{A5}), (\ref{A6}), and (\ref{A7}), the conditions $p_{1} =0$, $p_{2} =0$, and $p_{3} =0$ reperensts the solution above Curie temperature ($\theta>\theta_{C}$) as the material exhibits the paraelectric nature above $\theta_{C}$ and polarization vanishes in absence of electric field. This phase is referred as cubic phase, and the solution for cubic phase can be written as:
\begin{eqnarray}
Cubic & : & p_{1}=p_{2}=p_{3}=0,\label{A8}
\end{eqnarray}
BNT below $\theta_{C}$ exhibit tetragonal, orthorhombic and rhombohedral phase structure depending on the direction of polarization. In tetragonal phase, the direction of polarization is along (0,0,1) direction, therefore the solution for tetragonal phase is:
\begin{eqnarray}
Tetragonal & : & p_{1}=p_{2}=0,p_{3}\neq0, \quad \alpha_{1}\frac{\theta_{c}-\theta}{\theta_{c}}+2\alpha_{11}p_{3}^{2}+3\alpha_{111}p_{3}^{4}=0.\label{A9}
\end{eqnarray}
The direction of polarization in orthorhombic phase and rhombohedral phase is along (0,1,1) and (1,1,1) directions respectively, therefore the solution for orthorhombic phase and rhombohedral phase can be written as:
\begin{eqnarray}
Orthorombic & : & p_{1}=0, p_{2}=p_{3}\neq0, \quad \alpha_{1}\frac{\theta_{c}-\theta}{\theta_{c}}+(2\alpha_{11}+\alpha_{12}\frac{\theta_{c}-\theta}{\theta_{c}})p_{3}^{2}+3\alpha_{111}p_{3}^{4}=0,\label{A10}
\end{eqnarray}
\begin{eqnarray}
Rhombohedral & : & p_{1}=p_{2}=p_{3}\neq0, \quad \alpha_{1}\frac{\theta_{c}-\theta}{\theta_{c}}+2(\alpha_{11}+\alpha_{12}\frac{\theta_{c}-\theta}{\theta_{c}})p_{3}^{2}+3\alpha_{111}p_{3}^{4}=0.\label{A11}
\end{eqnarray}
On the basis of above solution, the Landau-Ginzburg-Devonshire free energy function for each phase can be written as:
\begin{eqnarray}
Cubic & : & W^{3DC}(\theta,\pmb{\overrightarrow{p}})=0,\label{A12}
\end{eqnarray}
\begin{eqnarray}
Tetragonal & : & W^{3DT}(\theta,\pmb{\overrightarrow{p}})=\alpha_{1}\frac{\theta_{c}-\theta}{\theta_{c}}p_{3}^{2}+\alpha_{11}p_{3}^{4}+\alpha_{111}p_{3}^{6},\label{A13}
\end{eqnarray}
\begin{eqnarray}
Orthorombic & : & W^{3DO}(\theta,\pmb{\overrightarrow{p}})=2\alpha_{1}\frac{\theta_{c}-\theta}{\theta_{c}}p_{3}^{2}+(2\alpha_{11}+\alpha_{12}\frac{\theta_{c}-\theta}{\theta_{c}})p_{3}^{4}+2\alpha_{111}p_{3}^{6},\label{A14}
\end{eqnarray}
\begin{eqnarray}
Rhombohedral & : & W^{3DR}(\theta,\pmb{\overrightarrow{p}})=3\alpha_{1}\frac{\theta_{c}-\theta}{\theta_{c}}p_{3}^{2}+3(\alpha_{11}+\alpha_{12}\frac{\theta_{c}-\theta}{\theta_{c}})p_{3}^{4}+3\alpha_{111}p_{3}^{6}.\label{A15}
\end{eqnarray}
\subsection{2D case}
For a two-dimensional problem along $x_{1}-x_{3}$ plane, the components of polarization vector are $p_{1}$ and $p_{3}$ only. Therefore, the minimization of $\phi$ with respect to $\pmb{\overrightarrow{p}})$ will also lead to minimization of $W^{2D}(\theta,\pmb{\overrightarrow{p}})$:
\begin{equation}
\phi_{\pmb{\overrightarrow{p}}} = W^{2D}_{\pmb{\overrightarrow{p}}}(\theta,\pmb{\overrightarrow{p}})-\pmb{\overrightarrow{E}} = 0. \label{A16}
\end{equation}
The solution of Eq. (\ref{A16}) will give us both component of electric field along $x_{1}-x_{3}$, which can be written as:
\begin{equation}
E_{1} = 2\alpha_{1}\frac{\theta_{c}-\theta}{\theta_{c}}p_{1}+4\alpha_{11}p_{1}^{3}+2\alpha_{12}\frac{\theta_{c}-\theta}{\theta_{c}}p_{1}p_{3}^{2}+6\alpha_{111}p_{1}^{5}, \label{A17}
\end{equation}
\begin{equation}
E_{3} = 2\alpha_{1}\frac{\theta_{c}-\theta}{\theta_{c}}p_{3}+4\alpha_{11}p_{3}^{3}+2\alpha_{12}\frac{\theta_{c}-\theta}{\theta_{c}}p_{3}p_{1}^{2}+6\alpha_{111}p_{3}^{5}. \label{A18}
\end{equation}
In the absence of electric field, the solution of Eqs. (\ref{A17}), and (\ref{A18}) can be written as:
\begin{equation}
p_{1} =0, \alpha_{1}\frac{\theta_{c}-\theta}{\theta_{c}}+2\alpha_{11}p_{1}^{2}+\alpha_{12}\frac{\theta_{c}-\theta}{\theta_{c}}p_{3}^{2}+3\alpha_{111}p_{1}^{4} = 0, \label{A19}
\end{equation}
\begin{equation}
p_{3} =0, \alpha_{1}\frac{\theta_{c}-\theta}{\theta_{c}}+2\alpha_{11}p_{3}^{2}+\alpha_{12}\frac{\theta_{c}-\theta}{\theta_{c}}p_{1}^{2}+3\alpha_{111}p_{3}^{4} = 0. \label{A20}
\end{equation}
Eqs. (\ref{A19}), and (\ref{A20}) are subjected to the solutions $p_{1} =0$, and $p_{3} =0$,  which represents the solution above the Curie temperature ($\theta>\theta_{C}$), when the material demonstrates its paraelectric nature and the polarization disappears in the absence of an electric field. This phase is known as the cubic phase, and the solution for this cubic phase may be expressed as:
\begin{eqnarray}
Cubic & : & p_{1}=p_{3}=0.\label{A21}
\end{eqnarray}
BNT exhibits tetragonal, orthorhombic, and rhombohedral phases below Curie temperature. For a two-dimensional plane, the tetragonal phase and orthorhombic phase exhibits the polarization in the (0,1) direction, therefore, the solution for these phases using Eq. (\ref{A20}) are represented as:
\begin{eqnarray}
Tetragonal & : & p_{1}=0,p_{3}\neq0, \quad \alpha_{1}\frac{\theta_{c}-\theta}{\theta_{c}}+2\alpha_{11}p_{3}^{2}+3\alpha_{111}p_{3}^{4}=0,\label{A22}
\end{eqnarray}
\begin{eqnarray}
Orthorombic & : & p_{1}=0,p_{3}\neq0, \quad \alpha_{1}\frac{\theta_{c}-\theta}{\theta_{c}}+2\alpha_{11}p_{3}^{2}+3\alpha_{111}p_{3}^{4}=0.\label{A23}
\end{eqnarray}
The direction of polarization in rhombohedral phase in $x_{1}-x_{3}$ plane is along the (1,1) direction, therefore the solution for rhombohedral phase is:
\begin{eqnarray}
Rhombohedral & : & p_{1}=p_{3}\neq0, \quad\alpha_{1}+(2\alpha_{11}+\alpha_{12}\frac{\theta_{c}-\theta}{\theta_{c}})p_{3}^{2}+3\alpha_{111}p_{3}^{4}=0.\label{A24}
\end{eqnarray}
The corresponding Landau-Ginzburg-Devonshire energy function for each
phase are as follows: 
\begin{eqnarray}
Cubic & : & W^{2DC}(\theta,\pmb{\overrightarrow{p}})=0,\label{A25}
\end{eqnarray}
\begin{eqnarray}
Tetragonal & : & W^{2DT}(\theta,\pmb{\overrightarrow{p}})=\alpha_{1}\frac{\theta_{c}-\theta}{\theta_{c}}p_{3}^{2}+\alpha_{11}p_{3}^{4}+\alpha_{111}p_{3}^{6},\label{A26}
\end{eqnarray}
\begin{eqnarray}
Orthorombic & : & W^{2DO}(\theta,\pmb{\overrightarrow{p}})=\alpha_{1}\frac{\theta_{c}-\theta}{\theta_{c}}p_{3}^{2}+\alpha_{11}p_{3}^{4}+\alpha_{111}p_{3}^{6},\label{A27}
\end{eqnarray}
\begin{eqnarray}
Rhombohedral & : & W^{2DR}(\theta,\pmb{\overrightarrow{p}})=2\alpha_{1}\frac{\theta_{c}-\theta}{\theta_{c}}p_{3}^{2}+(2\alpha_{11}+\alpha_{12}\frac{\theta_{c}-\theta}{\theta_{c}})p_{3}^{4}+2\alpha_{111}p_{3}^{6}.\label{A28}
\end{eqnarray}
\section{Calculation of Landau Coefficients}\label{B}
To determine the values of $\alpha_{1}$, $\alpha_{11}$, $\alpha_{12}$, and $\alpha_{111}$, we analyze the experimental data at the Curie point. Beyond this point, the phase exhibits cubic phase and zero polarization, therefore; the Landau-Ginzburg-Devonshire free
energy is zero beyond this point despite the large value of $\alpha_{1}$ as
$\theta>\theta_{c},$which is greater than $\theta_{0}$. However,
when the temperature is slightly lower than the Curie temperature, the variable $\alpha_{1}$ assumes a negative value. By keeping the
values of $\alpha_{11}$,
$\alpha_{12}$ and $\alpha_{111}$ constant, the Landau-Ginzburg-Devonshire
free energy polynomial function $W^{3D}(\theta,\pmb{\overrightarrow{p}})$ changes to the Eqs. (\ref{A13}), (\ref{A14}) and (\ref{A15}). Moreover, in the vicinity of the Curie temperature, the presence of an equilbrium between the tetragonal and cubic phases is seen, and  $W^{3D}(\theta,\pmb{\overrightarrow{p}})$ can be written as \citep{Wang2013,Devonshire1949}:
\begin{eqnarray}
W^{3D}(\theta,\pmb{\overrightarrow{p}}) & = &\alpha_{1c}p_{3c}^{2}+\alpha_{11}p_{3c}^{4}+\alpha_{111}p_{3c}^{6}= 0,\label{A29} 
\end{eqnarray}
where the subscript $c$ represents the values at Curie temperature. Now Eq. (\ref{A29}) can be written as:
\begin{eqnarray}
\alpha_{11}p_{3c}^{2}+\alpha_{111}p_{3c}^{4}= -\alpha_{1c},\label{A30} 
\end{eqnarray}
For Curie point rearranging the Eq. (\ref{A9}), we have:
\begin{eqnarray}
2\alpha_{11}p_{3c}^{2}+3\alpha_{111}p_{3c}^{4}= -\alpha_{1c}.\label{A31} 
\end{eqnarray}
Now by solving the Eqs. (\ref{A30}), and (\ref{A31}), the values of $\alpha_{11}$,
$\alpha_{111}$ are calculated as \citep{Wang2013,Devonshire1949}:
\begin{eqnarray}
\alpha_{11} & = & \frac{-2\alpha_{1c}}{p_{3c}^{2}}, \alpha_{111}  =  \frac{\alpha_{1c}}{p_{3c}^{4}}.\label{A32}
\end{eqnarray}
The variable $\alpha_{1}$ exhibits a strong correlation with the reciprocal of the dielectric susceptibility $(\chi)$, and can be determined $\alpha_{1}$
as follows \citep{Wang2013,Devonshire1949}:
\begin{eqnarray}
2\alpha_{1} & = & \frac{1}{\chi}=\frac{(\theta-\theta_{0})}{(\theta_{c}-\theta_{0})}\alpha_{1c}.\label{A33}
\end{eqnarray}
The value of $\alpha_{12}$ is considered proportional to $\alpha_{11}$ and can
be written as $\alpha_{12}=-a\alpha_{11}$, The determination of the value of $a$ involves fitting it to the Landau-Ginzburg-Devonshire free energy term at lower transition temperatures and subsequently comparing it to the experimental Landau-Ginzburg-Devonshire free energy at that specific point \citep{Wang2013,Devonshire1949}.
\section{Temperature dependence of spontaneous polarization}\label{C}
The remenant polarization on absence of electric field is defined as spontaneous polarization, which varies with the change of temperature as follows:
\subsection{3D case}
We get the temperature dependent spontaneous polarization $p_{s}(\theta)$ by solving Eqs. (\ref{A9}), (\ref{A10}), and (\ref{A11}) by substituting $p_{3}=p_{s}$ and keeping all other variables constant and can be given as follows:
\begin{eqnarray}
Tetragonal & : & p_{s}(\theta) =\sqrt{\frac{\sqrt{\alpha_{11}^{2}-3\alpha_{1}\alpha_{111}\frac{\theta_{c}-\theta}{\theta_{c}}}-\alpha_{11}}{3\alpha_{111}}},\label{A34}
\end{eqnarray}
\begin{eqnarray}
Orthorombic & : &  p_{s}(\theta) =\sqrt{\frac{\sqrt{(2\alpha_{11}+\alpha_{12}\frac{\theta_{c}-\theta}{\theta_{c}})^{2}-12\alpha_{1}\alpha_{111}\frac{\theta_{c}-\theta}{\theta_{c}}}-(2\alpha_{11}+\alpha_{12}\frac{\theta_{c}-\theta}{\theta_{c}})}{6\alpha_{111}}},\label{A35}
\end{eqnarray}
\begin{eqnarray}
Rhombohedral & : & p_{s}(\theta) =\sqrt{\frac{\sqrt{(\alpha_{11}+\alpha_{12}\frac{\theta_{c}-\theta}{\theta_{c}})^{2}-3\alpha_{1}\alpha_{111}\frac{\theta_{c}-\theta}{\theta_{c}}}-(\alpha_{11}+\alpha_{12}\frac{\theta_{c}-\theta}{\theta_{c}})}{3\alpha_{111}}},\label{A36}
\end{eqnarray}
\subsection{2D case}
\begin{figure} [!h]
\centering
{\includegraphics[scale=0.55]{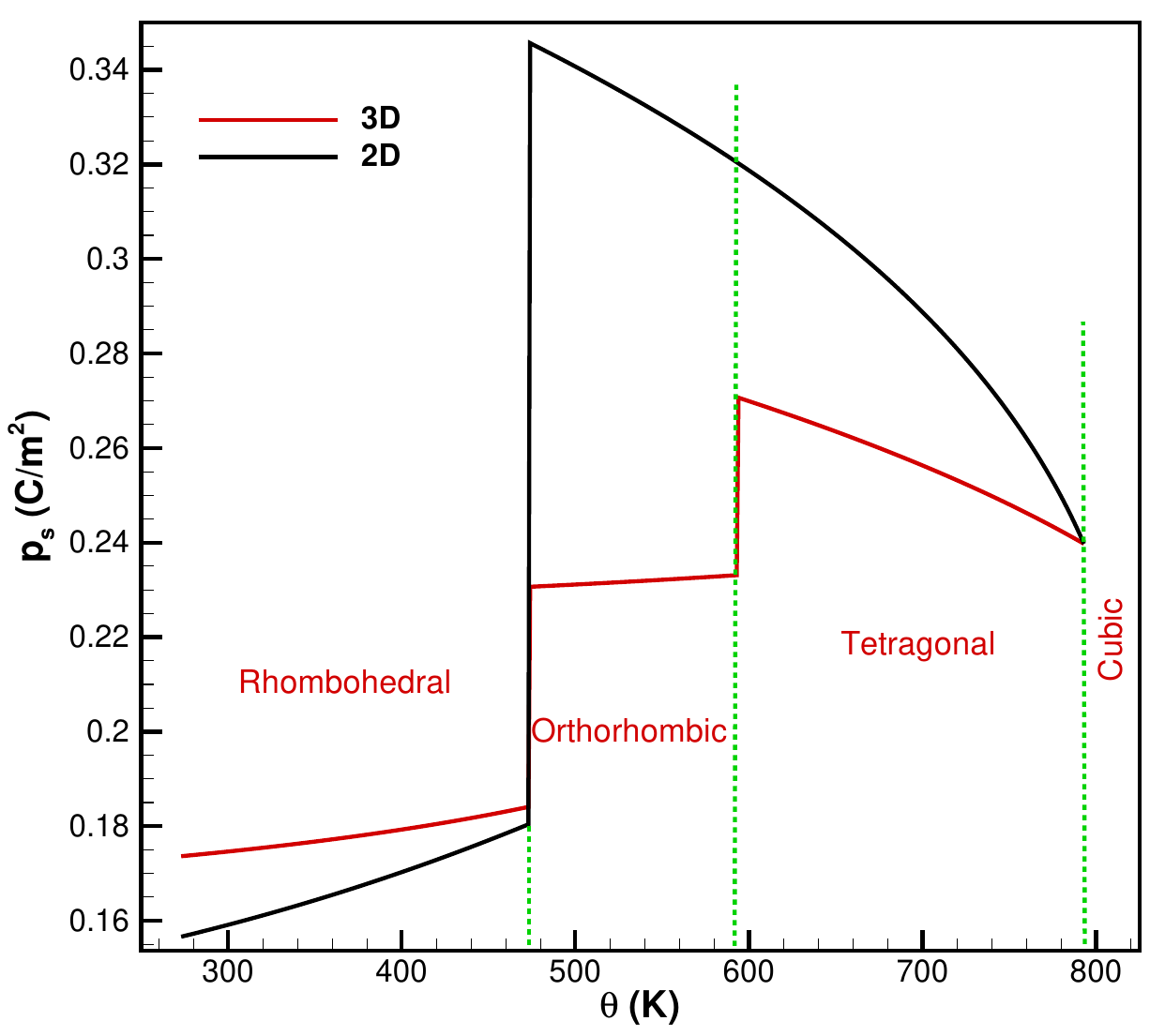}}
\caption { Variation of spantaneous polarization ($p_{s}$) with temperature ($\theta$)}
\label{fig:A1}
\end{figure}

We get the temperature dependent spontaneous polarization $p_{s}(\theta)$ by solving Eqs. (\ref{A22}), (\ref{A23}), and (\ref{A24}) by substituting $p_{3}=p_{s}$ and keeping all other variables constant and can be given as follows:
\begin{eqnarray}
Tetragonal ~and  ~Orthorombic& : & p_{s}(\theta) =\sqrt{\frac{\sqrt{\alpha_{11}^{2}-3\alpha_{1}\alpha_{111}\frac{\theta_{c}-\theta}{\theta_{c}}}-\alpha_{11}}{3\alpha_{111}}},\label{A37}
\end{eqnarray}
\begin{eqnarray}
Rhombohedral & : &  p_{s}(\theta) =\sqrt{\frac{\sqrt{(2\alpha_{11}+\alpha_{12}\frac{\theta_{c}-\theta}{\theta_{c}})^{2}-12\alpha_{1}\alpha_{111}\frac{\theta_{c}-\theta}{\theta_{c}}}-(2\alpha_{11}+\alpha_{12}\frac{\theta_{c}-\theta}{\theta_{c}})}{6\alpha_{111}}},\label{A38}
\end{eqnarray}
$p_{s}$ is plotted against $\theta$ for 2D and 3D case for comparision in Fig \ref{fig:A1}.
\section{Comparision of thermal evolution of spontaneous polarization for BNT and PZT}\label{D}

\begin{figure} [!h]
\centering
{\includegraphics[scale=0.55]{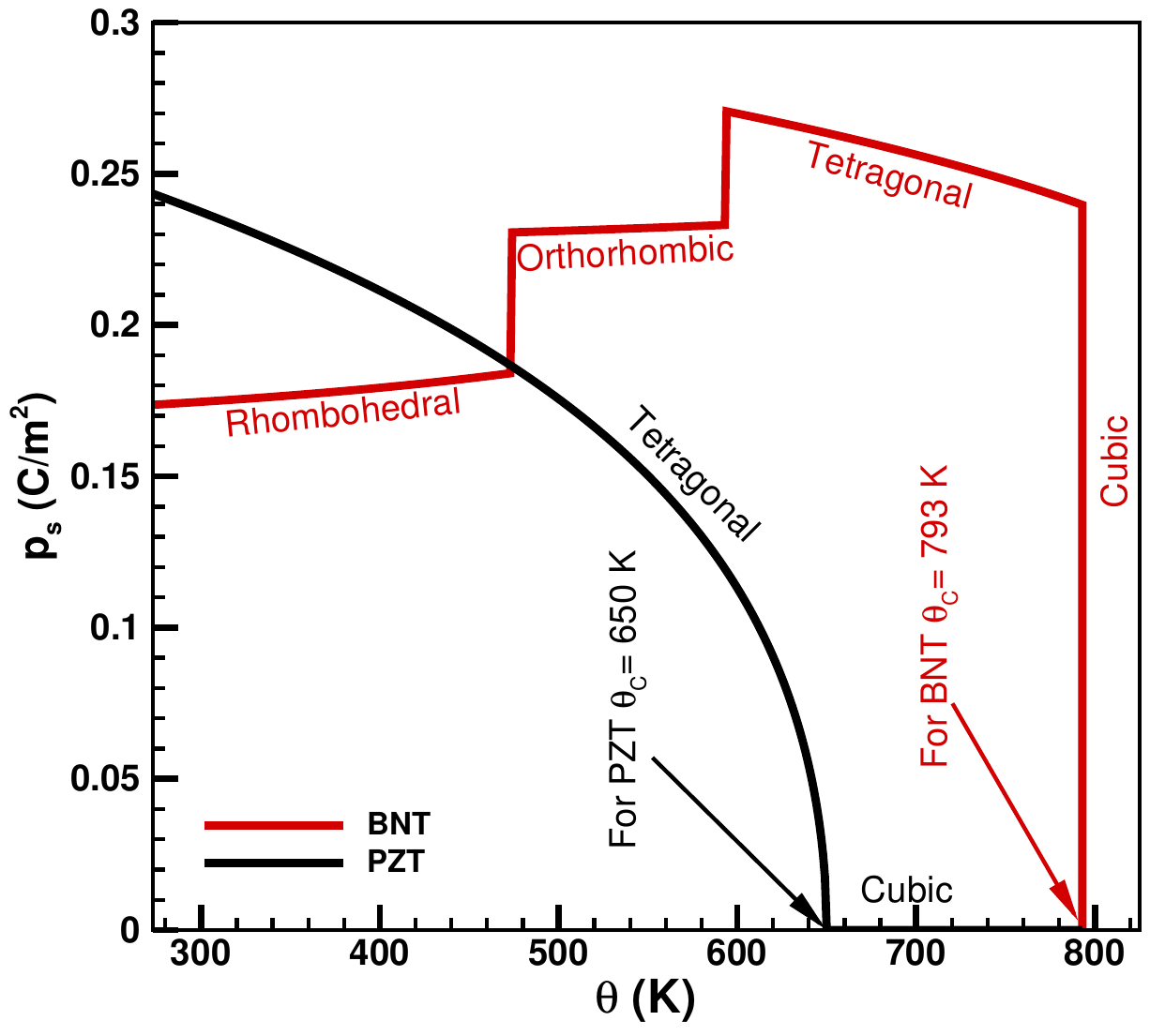}}
\caption { Comparision of variation of spantaneous polarization ($p_{s}$) with temperature ($\theta$) for BNT and PZT}
\label{fig:A2}
\end{figure}
The comarision of the variation of the sponatneous polarization with temperature for BNT and PZT are shown in Fig. \ref{fig:A2}. PZT 50/50 exhibits only tetragonal phase from reference temperature 300 $K$  to its Curie temperature which is 650 $K$, where spontaneous polarization decreases with temperature as shown by Eq. (\ref{A34})  and thereafter, it exhibits cubic phase where sponatneous varition is zero. However, BNT exhibits a complex phase structure and shows rhombohedral, orthorhombic and tetragonal phase from reference temperature to Curie temperature (793 $K$) and variation of sponatneous polarization with temperature as shown by Eqs. (\ref{A34}), (\ref{A35}), and (\ref{A36}) respectively. After Curie temperature, the spontaneous polarization is zero in cubic phase. 

\bibliographystyle{IEEEtranN}
\addcontentsline{toc}{section}{\refname}\bibliography{Phase_References}

\section*{\textemdash \textemdash \textemdash \textemdash \textemdash \textemdash \textemdash{}}
\end{document}